\documentclass[final,12pt,authoryear]{elsarticle}

\usepackage{amssymb}
\usepackage{lineno}
\usepackage{hyperref}
\usepackage{siunitx}
\usepackage{epstopdf}
\usepackage{xcolor}

\newcommand{\dz}[0]{$\delta z$}
\newcommand{\R}[1]{\textcolor{black}{#1}}
\newcommand{\RR}[1]{\textcolor{black}{#1}}

\journal{Ocean Modelling}

\begin{document}

\begin{frontmatter}

\title{Lagrangian surface signatures reveal upper-ocean vertical displacement conduits near oceanic density fronts}

\author[1]{H. M. Aravind}
\author[2]{Vicky Verma}
\author[2]{Sutanu Sarkar}
\author[3]{Mara A. Freilich}
\author[4]{Amala Mahadevan}
\author[5]{Patrick J. Haley}
\author[5]{Pierre F. J. Lermusiaux}
\author[1]{Michael R. Allshouse\corref{cor1}}
\cortext[cor1]{m.allshouse@northeastern.edu}

\affiliation[1]{organization={Mechanical and Industrial Engineering, Northeastern University},
            addressline={},
            city={Boston},
            postcode={02115},
            state={MA},
            country={USA}}
\affiliation[2]{organization={Mechanical and Aerospace Engineering, University of California San Diego},
            addressline={},
            city={La Jolla},
            postcode={92093},
            state={CA},
            country={USA}}
\affiliation[3]{organization={MIT–WHOI Joint Program in Oceanography},
            addressline={},
            city={Cambridge},
            postcode={02139},
            state={MA},
            country={USA}}
\affiliation[4]{organization={Woods Hole Oceanographic Institution},
            addressline={},
            city={Woods Hole},
            postcode={02543},
            state={MA},
            country={USA}}
\affiliation[5]{organization={Department of Mechanical Engineering, Massachusetts Institute of Technology},
            addressline={},
            city={Cambridge},
            postcode={02139},
            state={MA},
            country={USA}}

\begin{abstract}
\RR{Vertical transport in the ocean plays a critical role in the exchange of freshwater, heat, nutrients, and other biogeochemical tracers.
While there are situations where vertical fluxes are important, studying the vertical transport and displacement of material requires analysis over a finite interval of time.
One such example is the subduction of fluid from the mixed layer into the pycnocline, which is known to occur near density fronts.
Divergence has been used to estimate vertical velocities indicating that surface measurements, where observational data is most widely available, can be used to locate these vertical transport conduits.
We evaluate the correlation between surface signatures derived from Eulerian (horizontal divergence, density gradient, and vertical velocity) and Lagrangian (dilation rate and finite time Lyapunov exponent) metrics and vertical displacement conduits.
Two submesoscale resolving models of density fronts and a data-assimilative model of the western Mediterranean were analyzed.
The Lagrangian surface signatures locate significantly more of the strongest displacement features and the difference in the expected displacements relative to Eulerian ones increases with the length of the time interval considered.
Ensemble analysis of forecasts from the Mediterranean model demonstrates that the Lagrangian surface signatures can be used to identify regions of strongest downward vertical displacement even without knowledge of the true ocean state.}
\end{abstract}

\begin{graphicalabstract}
\end{graphicalabstract}

\begin{highlights} 
\item Surface Lagrangian metrics locate downward vertical displacement near density fronts

\item Method is robust for LES, process study model, and Mediterranean forecast models

\item Lagrangian target zones locate displacement that persists beyond time window analyzed

\item Ensemble forecast sufficient for locating displacement even without ``true" fields

\end{highlights}


\end{frontmatter}


\section{Introduction}

Vertical transport in the upper ocean is instrumental in surface mixing, nutrient supply in the biogeochemical cycle~\citep{denman1995biological,martin01,mahadevan16}, and the ocean energy budget~\citep{wunsch04,gnanadesikan05,siegelman20}.
It is also one of the mechanisms that govern the distribution of microplastics~\cite{reisser2015vertical} and other chemical pollutants~\citep{tanabe1983vertical,gonzalez2019vertical} in the ocean.
Observational investigation of vertical transport, however, is considerably more difficult than studying horizontal transport because vertical velocities in the ocean $\big(O(\SI{e-4}{\metre\per\second})\big)$ are orders of magnitude smaller than the horizontal components $\big(O(\SI{0.1}{\metre\per\second})\big)$~\citep{mahadevan16,mahadevan20}.
Additionally, the vertical velocity magnitudes are near the noise-limit for most acoustic Doppler current profilers~\citep{fischer93}, making direct measurements difficult.
The combined impact of waves, eddies, and wind-driven Ekman dynamics result in a highly variable vertical velocity field where the influence of any one phenomenon is hard to distinguish.

One quantity of interest when studying vertical transport is the time-integrated vertical displacement (i.e. change in depth) of fluid-particle trajectories over a time interval of hours to days.
These measurements of vertical displacements are on the order of tens to hundreds of meters over fluid-particle trajectories that extend kilometers in the horizontal direction.
\RR{Even when downward displacement of fluid is observed in the ocean, only a fraction undergoes ``subduction,'' which is the transfer of fluid from the surface mixed layer into the pycnocline and stratified interior~\citep{spall1995frontogenesis,marshall1997subduction,williams2019ocean,mahadevan20}}.
Oceanographers have proposed the existence of three-dimensional coherent pathways at sites of subduction that act as conduits for exchange from the surface to the interior~\citep{calypso19,mahadevan20}.
For short time intervals, particles inside vertical transport conduits initially near the surface can undergo downward displacement but remain in the mixed layer while those that start deeper may undergo subduction by leaving the mixed layer.
Recent studies have shown that these sites are closely linked to submesoscale ($O(\SI{1}{\kilo\metre})$ horizontal length scale) features~\citep{mcwilliams2016submesoscale} such as density fronts, and are not randomly distributed~\citep{ruiz20}.

Numerical studies based on high resolution ocean models demonstrate how submesoscale dynamics \R{near oceanic density fronts} dominate vertical transport in the ocean and that Ekman pumping resulting from wind forcing on the surface is insufficient to fully account for the observed vertical transport signatures~\citep{mahadevan06,giordani06,thomas08,nagai2015dominant}.
Field studies guided by insights from numerical studies of submesoscale eddies~\citep{omand15} and oceanic density fronts~\citep{pascual17,mahadevan20} have confirmed the importance of submesoscale processes in dictating ocean vertical transport.
More recent numerical studies have suggested the existence of filaments of concentrated vorticity that are $O(\SI{0.1}{km})$ wide along the perimeter of submesoscale eddies that serve as coherent pathways for vertical motions in the ocean~\citep{mahadevan06filaments,omand15,freilich19,verma2021lagrangian}.
Although numerical simulations using ocean models can provide insight into vertical transport mechanisms by generating three-dimensional velocity fields, a combination of ocean modeling and observational field experiments~\citep{lermusiaux2006progress} is required to understand and verify the various mechanisms that result in enhanced vertical velocities and the underlying pathways~\citep{omand15,dasaro18} of vertical transport.
Experimental investigations of these coherent pathways require precise targeting of the conduits that may be located along submesoscale vorticity filaments near oceanic density fronts.

In order to locate vertical transport conduits during an observational study, the method should rely on ocean surface signatures where measurements are most prevalent.
To identify regions near the ocean surface where strong vertical transport is anticipated, surface fields from ocean forecasts and satellite altimetry are used~\citep{mahadevan20}.
Ideally, a vertical velocity field could be used to locate regions of strong instantaneous \R{downwelling}, but reliable estimates are hard to obtain due to being orders of magnitude smaller than the horizontal components.
Alternatively, computation of the negative horizontal divergence of the surface horizontal velocity field can highlight regions of instantaneous surface convergence, which correspond to downwelling at that instant.
During field experiments, the divergence can be estimated using trajectories of drifter swarms~\citep{molinari75,berta16,dasaro18,berta20,tarry2021frontal}.
Recent numerical and observational studies have investigated the link between divergence and enhanced vertical motion near oceanic density fronts~\citep{barkan19}.
The instantaneous location of these fronts, which are known to feature enhanced vertical velocities, can be identified using the horizontal density gradient on the surface.
While some of these Eulerian metrics are measurable during field experiments, they are only capable of revealing the instantaneous state of the ocean like \RR{an instantaneous ``subduction rate,'' which is often defined as the flux of fluid through a depth surface~\citep{marshall1993inferring,qu2009north}}.
Hence, their ability to predict vertical displacement, an intrinsically Lagrangian phenomenon that occurs over a finite time interval, may be limited.

A Lagrangian analysis is more indicative of cumulative vertical transport, but the predictive value of a Lagrangian analysis of real-time data to identify sites of deeper displacement beyond the time interval of analysis is unknown.
Moreover, individual trajectories computed from model forecasts are highly sensitive to noise and model error, so it may not provide robust prediction of where vertical transport may occur in the real ocean.
Lagrangian metrics such as the finite-time Lyapunov exponent (FTLE)~\citep{haller01} and dilation rate~\citep{huntley15}, which are robust to noise~\citep{haller02}, can be used to identify regions of surface convergence.
These metrics have been widely employed in the calculation of Lagrangian coherent structures~\citep{allshouse15,haller15} to identify transport barriers in fluid flows.
Two-dimensional coherent structure calculations have been used to identify two-dimensional transport features~\citep{olascoaga06,d2009comparison,hernandez-carrasco18} and link them to surface circulation and local mixing~\citep{choi2017submesoscale}.
These structures and their three-dimensional extensions, however, have yet to be used to target vertical transport through coherent conduits in the upper ocean.
Performing a three dimensional coherent structure analysis may potentially reveal the structure of these conduits, but it is not feasible due to the lack of accurate three dimensional oceanic fields.
Furthermore, leveraging the available drifter data requires performing an analysis that restricts trajectory advection to the two dimensional surface.
In a general three dimensional system, the analysis of a two dimensional slice of the domain provides useful information about the full three dimensional transport only in unique conditions~\citep{sulman13}.
Fortunately, in an oceanic setting, where large scale motions are predominantly horizontal, it is reasonable to consider the two-dimensional surface signatures as representative of the weakly three dimensional processes in the upper ocean.
In fact, field experiments that investigate subduction in the ocean look for some surface signatures, like convergence of drifter trajectories~\citep{dasaro18,mahadevan20}.

\R{Because the available measurements are drawn from the ocean surface, the influence of these surface converging zones will be most relevant for vertical displacements and subduction of particles that are initially near the surface.
Two challenges with this initial placement are that these particles will be exposed to the weak vertical velocities near the surface and will have to be transported through the entire mixed layer to subduct into the ocean interior.
As a result, some of the vertical displacement observed may not necessarily mean that a particular particle has undergone subduction.
However, the particles in the strongest vertical displacement conduits can potentially leave the mixed layer if a longer time interval is considered or if they are initialized deeper in the water column.
Hence, our analysis focuses on linking the surface Lagrangian and Eulerian metrics to regions of significant vertical displacement.}

To test the ability of Lagrangian and Eulerian surface metrics to \R{locate regions of strong downward displacement of fluid}, we analyze the method applied to three ocean models ranging in scale from kilometers to hundreds of kilometers.
The smallest model considered is a large eddy simulation model that captures the detailed dynamics of density front filamentation and how the turbulence impacts the local transport~\cite{verma19}.
Next, a process study ocean model considers the evolution of a zonally oriented density front and its interaction with submesoscale eddies and vorticity filaments, with initial conditions constructed using observed temperature and salinity measurements in the Mediterranean~\citep{mahadevan96a,mahadevan96b,freilich19,freilich21,PSOM}.
Finally, a data-assimilative real-time operational model of the Mediterranean that comprises multiple density fronts that co-exist with other flow features like basin-wide gyres is considered, which provides the most realistic setting for field experiments~\citep{haley10,haley15,calypso19}.
Each of these models feature a density front, contain submesoscale features and utilize markedly different underlying assumptions.
\R{The analyses of \cite{verma2021lagrangian} and \cite{freilich21} have presented evidence of coherent downward transport in the turbulence resolving and process study models and, additionally, \cite{freilich21} find subduction to below the mixed layer.}
By applying our method to these models of different scale and physics, we are able to evaluate the robustness of applicability of Lagrangian and Eulerian metrics.

In this paper, we propose the use of two Lagrangian metrics computed using surface horizontal velocities to predict regions of high sub-surface \R{downward displacement of fluid} in the ocean.
The regions where these metrics are strongest are considered target zones for downward displacement, and the efficacy of these target zones in locating large vertical displacements is evaluated using skill metrics that quantify how likely they are to predict regions of \R{large displacements} and the fraction of such regions identified by the target zones.
The correspondence between the surface target zones and sub-surface \R{\dz\ fields computed for particles} initialized at different depths below the surface is then studied to evaluate the range of applicability of the target zones.
A practical concern when performing a Lagrangian analysis is the time interval of applicability of results.
Hence, we investigate the effect of varying the time interval of Lagrangian analysis, and the ability of the chosen Lagrangian metrics to predict \R{large displacements} occurring beyond that interval.
Finally, since model forecasts do not capture the true state of the ocean~\citep{mcwilliams2007irreducible}, the ability of target zones identified across an ensemble of forecasts to locate \R{regions of strong downward displacement} in a separate realization is evaluated with a data-assimilative operational model of the western Mediterranean.

\section{Methods}
\subsection{Vertical Displacement, Surface Metrics and Target Zones.}
Conduits of vertical transport can be located by identifying regions with fluid trajectories that show large vertical displacements.
Given a velocity field $\mathbf{u}=(u,v,w)$ specified at $\mathbf{x}=(x,y,z)$, fluid trajectories are solutions to
\begin{equation} \label{eq:trajectories}
    \frac{\mathrm{d}\mathbf{x}}{\mathrm{d}t} = \mathbf{u}(\mathbf{x},t).
\end{equation}
For each of the three models, particles are seeded uniformly on a horizontal plane at $z(t=t_0)$ below the surface and then advected by solving equation~(\ref{eq:trajectories}) using a fourth order Runge-Kutta scheme.
\R{Velocities for particle advection are obtained from model fields using a cubic Lagrange polynomial for temporal interpolation, followed by a bilinear interpolation in space.}
The \R{downward vertical displacement} (\dz) undergone by particles seeded at depth $z(t=t_0)$ is then computed as
\begin{equation}
    \delta z = z(t_0) - z(t_0+T_d),
\end{equation}
where the displacement time $T_d$ is the time over which advection is performed.
In our study, a $\SI{12}{\hour}$ interval has been used for the two larger-scale models, whereas a smaller $T_d = \SI{6}{\hour}$ is used for the turbulence resolving model so that the non-dimensional value $T_d/(2\pi/f)$ remains approximately 0.5 across the three models.

Next, we compute Eulerian and Lagrangian metrics that can potentially identify regions of high \dz\ using only surface information.
The most intuitive measure of vertical motions, given a velocity field, is the vertical velocity $w$.
Because $w<0$ corresponds to downwelling, the negative vertical velocity ($-w$) at time $t_0$ is the first metric.
It is important to note that although $w$ highlights regions of strongest vertical motion at an instant, the Eulerian field need not be representative of vertical displacement that occurs over a finite time interval.
Another drawback of this metric is that vertical velocities are often difficult to measure in the ocean and model accurately.
The horizontal velocities, on the other hand, are measurable, and incompressibility dictates that convergence on the ocean surface corresponds to downwelling.
Therefore, the negative horizontal divergence of surface velocity ($-\nabla_H \cdot \mathbf{u}$), positive for downwelling, is the second Eulerian metric.
Physically, surface convergence is enhanced at submesoscale density fronts in the ocean, which are known to be sites of intense vertical motions~\citep{mahadevan06}.
Hence, the magnitude of the horizontal gradient of surface density ($|\nabla_H \rho|$), which can be measured observationally~\citep{pascual17,rudnick07,ullman14}, is the final Eulerian metric.
All of these Eulerian metrics are computed only at the initial instant, $t_0$.
The inability to obtain accurate measurements of $w$ prompts the usage of the other Eulerian metrics for practical purposes.
In this study, however, the Eulerian metric $w$ is available from the models and is used as a barometer to compare the other metrics with.
In all three models considered, $w$ is computed on the horizontal layer just below the ocean surface.
This can impact our results, especially in the process study model, as will be discussed later.

To better account for the Lagrangian nature of vertical displacement, two Lagrangian metrics are computed on the ocean surface: FTLE and dilation rate.
Solving equation~(\ref{eq:trajectories}) for $x$ and $y$ at $z=0$, using only the horizontal velocity and setting $w=0$, allows us to define the two-dimensional flow map, $\mathrm{F}_{t_0}^{t_0+T_L}(\mathbf{x}_0):=\mathbf{x}(t_0+T_L;\mathbf{x}_0,t_0)$, that maps particles seeded at $\mathbf{x}_0 = (x_0,y_0,0)$ at time $t_0$ to their final positions at $t_0+T_L$.
Here, $T_L$ is the length of the time interval over which the Lagrangian metrics are calculated, and this may or may not be the same as the displacement time, $T_d$, depending on the time over which accurate surface velocity fields are available.
The flow map gradient, $\nabla \mathrm{F}_{t_0}^{t_0+T_L}(\mathbf{x}_0)$, is then used to compute the right Cauchy-Green strain tensor $\mathrm{C}=[\nabla \mathrm{F}_{t_0}^{t_0+T_L}(\mathbf{x}_0)]^T[\nabla \mathrm{F}_{t_0}^{t_0+T_L}(\mathbf{x}_0)]$.
Given the two eigenvalues of $\mathrm{C}$, $0<\lambda_1<\lambda_2$, the FTLE ($\sigma$) at time $t_0$ is
\begin{equation}
    \sigma = \frac{\log{\lambda_2}}{2T_L},
\end{equation}
and the dilation rate ($\Delta$) at $t_0$ is
\begin{equation}
    \Delta = \frac{\log{\lambda_1\lambda_2}}{2T_L}.
\end{equation}
The dilation rate is the Lagrangian time-average of the Eulerian divergence ($\nabla_H \cdot \mathbf{u}$) along the trajectory~\citep{huntley15}.
It is worth noting that the Lagrangian metrics computed along particle trajectories differ from metrics obtained by averaging at a fixed location in space.
Another potential metric that could be considered is the instantaneous Lyapunov exponent (iLE)~\citep{nolan20}, which is the Eulerian infinitesimal time limit of FTLE.
This metric has been employed for the detection of objective Eulerian coherent structures~\citep{serra16}, but its ability to locate regions of strong \dz\ is similar to the other Eulerian metrics (Figure~S1, Supporting Information).

When the Lagrangian metrics are computed forward in time from $t_0$ to $t_0+T_L$, the two-dimensional FTLE and dilation rate fields obtained at the initial time $t_0$ physically represent the amount of maximal directional expansion and area expansion rates over the time interval, respectively.
Since we are interested in surface convergence instead of expansion, these quantities need to be computed starting at the end of the time interval $t_0+T_L$ and moving backwards in time to the beginning of the interval $t_0$~\citep{haller15}.
A backward-in-time calculation, however, yields Lagrangian fields at $t_0+T_L$, which do not correspond to the \dz\ field at $t_0$.
To make the direct comparison possible, we map the Lagrangian metric fields to their position at $t_0$.
\RR{Additionally, result independence from interpolation and advection parameters was ensured for all numerically extracted particle trajectories, and hence the results are comparable to those from online calculations performed during model runs~\cite{van2018lagrangian}.
}

The five metrics discussed above are chosen such that the regions with the largest values of these metrics should correspond to regions of greatest downwelling.
Hence, the target zone (TZ) for each metric is defined as the region with values in the \R{99th percentile of the respective Eulerian or Lagrangian metric computed on the ocean surface.
As a result, each target zone covers only 1\% of the entire domain.}
The choice of 1\% is based on our study varying the threshold (presented in Figure~S2, Supporting Information), such that the TZs identify only the strongest \dz\ regions, while being large enough to yield regions that are not trivially small.
The choice of the target threshold need not be universal; however, in the cases studied here, $1\%$ was found to be ideal.
The TZs can also be used to predict regions of strongest \dz\ when computed using model forecasts, as discussed later.
To quantify the ability of each metric's TZ to identify regions of strong \R{downward vertical displacement}, we investigate the extent of overlap between the TZ derived from the metric on the ocean surface and the regions of strongest \dz\  below the surface.

\subsection{Numerical Models Analysed.}
\label{subsec:methods_2}
\begin{figure}[t]
    \centering
    \includegraphics[width=\linewidth]{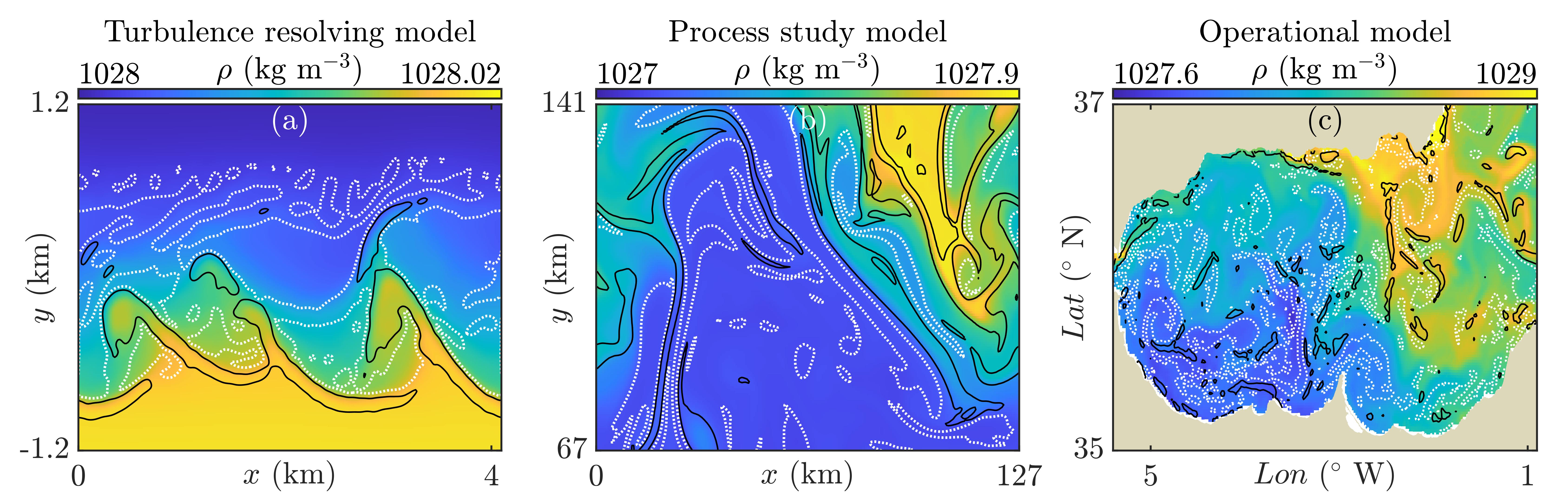}
    \caption{Models analysed. Surface density fields at the initial time ($t=t_0$) for (a) the turbulence resolving model of a front, (b) process study model of a front and (c) operational model of the western Mediterranean.
    Density fields are plotted in colour, and the gradients indicate locations of density fronts.
    Contours of the vertical vorticity non-dimensionalised by the planetary vorticity ($\omega_z/f$), which is indicative of the local Rossby number, are overlaid.
    The negative contours (anti-cyclonic vorticity, dashed white) correspond to $\omega_z/f$ values of (a) -0.5, (b) -0.5 and (c) -0.4, and the positive contours (cyclonic vorticity, solid black) to (a) 1, (b) 0.5 and (c) 0.8.
    The domains analyzed in (a) and (b) focus on the density front and do not span the entire numerical domain.}
    \label{fig:1}
\end{figure}

Three different ocean models are analysed to evaluate the robustness of the target zones capacity to identify sites of \R{strongest downward vertical displacement}.
\R{Given the connection between submesoscale features and subduction, we consider three models that each produce submesoscale features by accounting for different flow physics at spatial scales spanning two orders of magnitude.}
Figure~\ref{fig:1} presents the surface density field at $t_0$, the time at which we start our analysis.
The horizontal gradient in the density field highlights the location and structure of the density front in each system.
Contours of the non-dimensional vertical vorticity are plotted to highlight the regions of submesoscale fronts and eddies, which are related to regions of subduction~\citep{mahadevan06}.

The first model analysed is a large eddy simulation of an oceanic density front under the influence of rotation by solving the non-hydrostatic Navier-Stokes equations under the Boussinesq approximation~\citep{verma19}.
The domain for the turbulence resolving simulation spanned $\SI{4}{\kilo\metre} \times \SI{6}{\kilo\metre}$ in the horizontal, was $\SI{130}{\metre}$ deep, and had a resolution of \SI{2}{\metre} in all directions.
The initial three-dimensional temperature profile features a \SI{1.2}{\kilo\metre} wide east-west density front confined in a surface layer with depth \SI{50}{\metre}.
The initial velocity profile is obtained by integrating the density field to satisfy the thermal wind balance.
The specified set of initial conditions establishes a density front which is geostrophically balanced by a surface jet along the front.
Periodic boundary conditions are used along the front, and free-slip (velocity) and no-flux (temperature) conditions are used on the surface and cross-front boundaries.
A constant heat flux and free-slip condition for velocity are used on the bottom boundary.

This high resolution model captures the impact that turbulence has on the submesoscale eddies and the resulting vertical transport.
Since only submesoscale fluid motions play the dominant role in sustained downward transport~\citep{verma2021lagrangian} and the turbulence introduced features at length scales below the features we are interested in, we present our analysis on filtered submesoscale fields that remove the fine-scale turbulent fluctuations from the model output without removing the larger scale impact of modeling turbulence.
Corresponding analysis of the unfiltered fields over a shorter displacement time is presented in Figure~S3 (Supporting Information) and  yields qualitatively similar results to the filtered data.
The gradient in the submesoscale density field and the vorticity contours at $t_0$ (Figure~\ref{fig:1}a) reveal a density front deformed by baroclinic instability, resulting in the formation of submesoscale eddies and vorticity filaments along the front.
Modelled submesoscale vertical velocities are as large as $\SI{3.5E-3}{\metre\per\second}$ near the vorticity filaments and the periphery of the submesoscale vortices, where \R{downward displacement} of up to $\R{\delta z_{max}=}\sim\SI{35}{\metre}$ is observed over $\SI{6}{\hour}$.
This corresponds to a \R{vertical displacement rate} of $\R{\delta z_{max}/T_d=}\SI{1.6E-3}{\metre\per\second}$, which is an order of magnitude larger than vertical velocities observed in large-scale numerical models of the ocean.
The displacement time of $T_d = \SI{6}{\hour}$ corresponds to a non-dimensional value $T_d/(2\pi/f) = 0.48$, where $f = \SI{1.4e-4}{\per\second}$ is the Coriolis parameter for the model.
\R{The model outputs were saved approximately every \SI{400}{\second}, and three-dimensional particle advection was performed over $T_d = \SI{6}{\hour}$ using a time-step of \SI{60}{\second}.
To obtain converged two-dimensional trajectories over $T_L = T_d = \SI{6}{\hour}$ for computing the Lagrangian metrics using the horizontally divergent two-dimensional surface velocity field, however, a smaller time-step of \SI{10}{\second} was required.
Due to the strong convergence on the surface, analysis of two-dimensional trajectories computed over time intervals of \SI{12}{\hour} or more revealed non-negligible accumulation of numerical error even for smaller time-steps at the strongest converging regions.
Hence, we use a non-dimensional time of $T_d/(2\pi/f) \approx 0.5$ (which corresponds to $T_d = \SI{6}{\hour}$ for the turbulence resolving model) for all three models analyzed.}

Next, we analyse the process study ocean model~\citep{mahadevan96a,mahadevan96b,freilich19,freilich21,PSOM} initialised with a geostrophically balanced front oriented zonally in a periodic channel.
The model solves the non-hydrostatic Navier-Stokes equations in a $\SI{128}{\kilo\metre} \times \SI{206}{\kilo\metre} \times \SI{1}{\kilo\metre}$ domain, at a horizontal resolution of $\sim\SI{500}{\metre}$.
The domain has periodic boundary conditions in the east-west direction and free-slip conditions at the bottom, north, and south boundaries.
The initial temperature and salinity is constructed using observed Western Mediterranean thermocline structure with \SIrange[range-phrase = --]{50}{70}{\metre} mixed layer, and the model is cooled uniformly.
The temperature gradient on the surface results in an initial density front that runs in the east-west direction, and the initial flow is in thermal wind balance, without any wind forcing on the surface.
Subgridscale mixing is parameterized with a \SI{1}{\square\meter\per\second} horizontal diffusivity and \SI{1e-4}{\square\meter\per\second} vertical diffusivity.

Owing to its spatial resolution of $\SI{500}{\meter}$, the model does not capture the sub-kilometer-scale turbulence observed in the previous model, but it does simulate the baroclinic instability with submesoscale physics.
The density field at $t_0$ (Figure~\ref{fig:1}b) presents the evolved structure of the density front.
Contours of non-dimensional vorticity indicate the location of submesoscale eddies and vorticity filaments, which are observed on either side of the front.
The strongest vertical velocities observed in this model ($\sim\SI{4.5e-4}{\metre\per\second}$) are smaller than the turbulence resolving model, and hence, particles seeded near the surface are not displaced more than $\SI{8}{\metre}$ downward over $T_d = \SI{12}{\hour}$ --- here, $T_d/(2\pi/f) = 0.47$, where the Coriolis parameter $f = \SI{0.7e-4}{\per\second}$.
The vertical velocities in this model are smaller than the turbulence resolving model due to the large horizontal diffusivity, and the boundary layer turbulence parameterization.
\R{Both the two-dimensional and three-dimensional particle trajectory advection for this model were performed from the 3-hourly model output using a time-step of \SI{60}{\second} to obtain converged trajectories.}

The third model considered is a data-assimilative real-time operational model of the western Mediterranean~\citep{haley10,haley15,calypso19}.
The simulations were performed using the MIT Multidisciplinary Simulation, Estimation and Assimilation Systems (MIT-MSEAS) primitive equation ocean modeling system~\citep{haley10,haley15}.
The computational domain spanned approximately $\SI{430}{\kilo\metre}\times\SI{267}{\kilo\metre}$ in the horizontal at a resolution of $\sim\SI{500}{\metre}$, with 70 optimised terrain-following vertical levels based on the \verb|SRTM15+| bathymetry\citep{tozer19}.
Surface forcing for the simulations was based on the atmospheric fluxes from the $1/4^\circ$ NCEP Global Forecast System product and tidal forcing from the high-resolution \verb|TPXO8-atlas|\citep{egbert02}.
Initial conditions were obtained from three different models: $1/12^\circ$ Hybrid Coordinate Ocean Model, $1/24^\circ$ Copernicus Marine Environment Monitoring Service, and $\sim1/50^\circ$ Western Mediterranean Operational forecasting system.
These were corrected using observational data from Argo floats and, whenever possible, moorings.
The downscaled forecasts of the western Mediterranean were used as central forecasts for the ensemble forecasting method using the Error Subspace Statistical Estimation schemes~\citep{lermusiaux02}.
\R{The initial 3D perturbations were constructed from a combination of vertical empirical orthogonal functions of historical March CTD data along with an eigendecomposition of a horizontal correlation matrix computed using a 12.5km decay scale and a 31.25km zero-crossing.
A 3D primitive equation balance was applied to the perturbed velocities.}

The dense, salty water in the Mediterranean meets a jet of lighter water from the Atlantic entering through the Strait of Gibraltar, resulting in the formation of strong density fronts and coherent basin-scale gyres.
The gradient in the density field in Figure~\ref{fig:1}c identifies the location of these fronts on 12 Apr 2019 12:00:00, and the contours of non-dimensional vorticity roughly align with the location of the mesoscale gyres.
The gyres are not stationary, nor are they constantly present \cite{heburn1990variations, vargas2002seasonal, pascual17}.
A reference value of the Coriolis parameter for this region is $f = \SI{0.86e-4}{\per\second}$.
Vertical velocities of up to $\sim\SI{5e-3}{\metre\per\second}$ are observed near the surface, and \R{downward displacement} of up to $\R{\delta z_{max}=}\SI{130}{\metre}$ is observed over $T_d = \SI{12}{\hour}$, which corresponds to a non-dimensional displacement time of $T_d/(2\pi/f) = 0.59$.
The corresponding \R{vertical displacement rate} ($\R{\delta z_{max}/T_d=} \SI{3e-3}{\metre\per\second}$) is indicative of regions of strong vertical transport in the Mediterranean, owing to intense vertical motions near the fronts.
\R{Converged particle trajectories in the operational model were computed from 3-hourly model outputs using a time-step of \SI{60}{\second}.}

\section{Results}
We begin by analyzing the turbulence resolving model, which provides the highest fidelity model of the dynamics near a density front.
The \R{\dz\ field}, the five surface metrics and their corresponding target zones, computed for the turbulence resolving model, are presented in \S~\ref{subsec:results_1}.
The efficacy of TZs to identify regions of strongest \R{downward vertical displacement} is then quantified using various skill metrics for each of the models in \S~\ref{subsec:results_2}.
To understand the vertical length scale over which our results may be valid in a realistic setting, the effect of varying the seeding depth for \R{\dz}\ measurements is investigated for the operational model in \S~\ref{subsec:results_3}.
Then practical concerns of the impact of time intervals and analyzing ensemble simulations are investigated in \S~\ref{subsec:results_4} and \S~\ref{subsec:results_5}, respectively.

\subsection{Analysis of the Turbulence Resolving Model}
\label{subsec:results_1}

\begin{figure}[t]
    \centering
    \includegraphics[width=\linewidth]{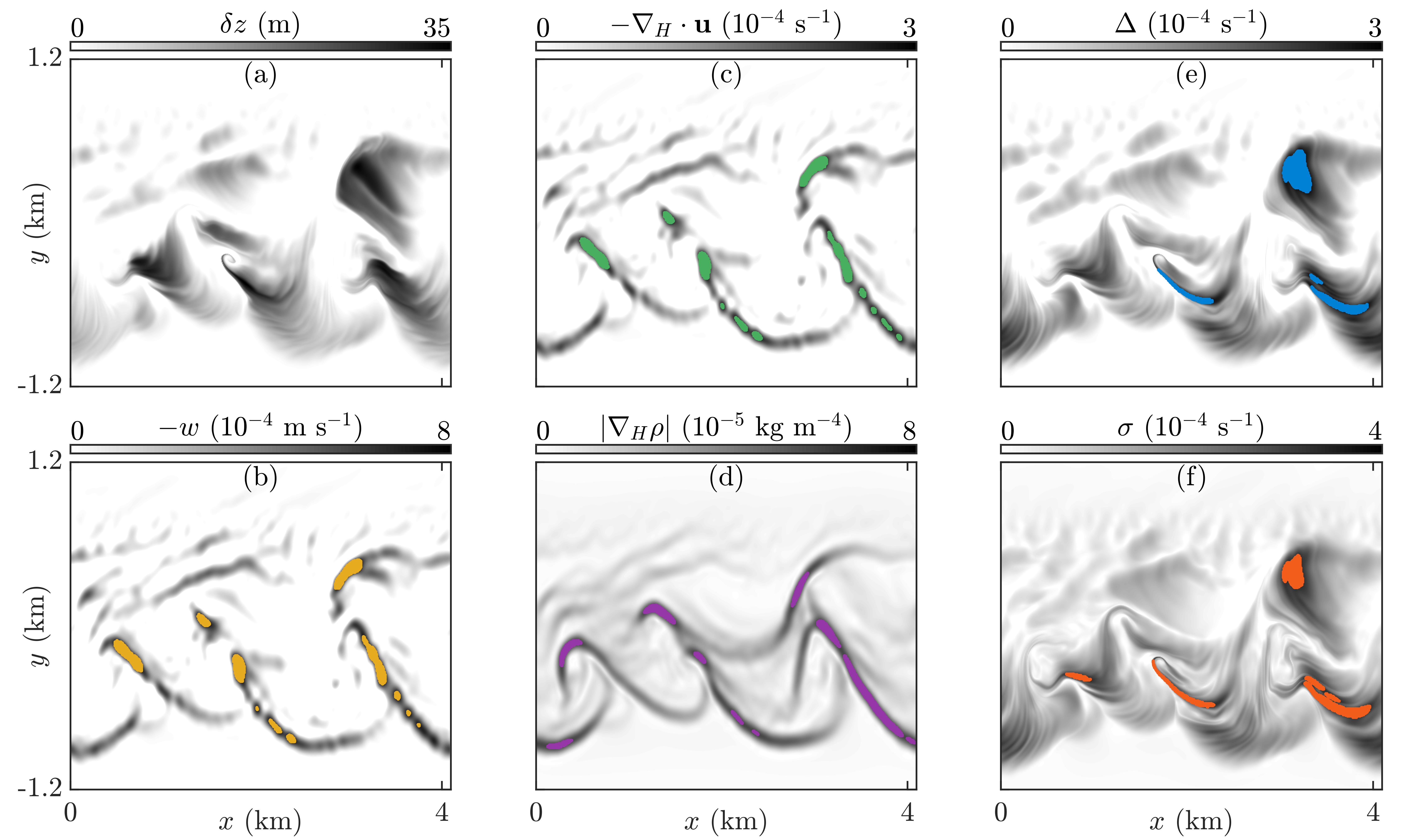}
    \caption{Downward vertical displacement below the surface, surface metrics and corresponding target zones for the turbulence resolving model of a density front. (a) \R{The \dz\ field} for particles released at a depth of \SI{4}{\metre}, computed over the interval $[t_0,t_0+T_d]$ and plotted at $t_0$. Eulerian surface target zones at $t_0$: (b) negative vertical velocity ($-w$), (c) negative horizontal divergence of velocity ($-\nabla_H \cdot \mathbf{u}$), and (d) horizontal density gradient magnitude ($|\nabla_H \rho|$). Lagrangian surface target zones, computed over the interval $[t_0,t_0+T_L]$ and plotted at $t_0$: (e) dilation rate ($\Delta$) and (f) FTLE ($\sigma$). The target zones for each metric are overlaid in color in panels (b)-(f). Note that, for the turbulence resolving model, $T_d=T_L=\SI{6}{\hour}$.}
    \label{fig:2}
\end{figure}

In this section, we discuss in detail the \R{\dz\ field}, the five surface metrics and their corresponding target zones computed for the turbulence resolving model of a density front.
The \R{\dz\ field} for particles released at $z(t_0) = -\SI{4}{\metre}$ is calculated over a displacement time $T_d = \SI{6}{\hour}$ and is presented in Figure~\ref{fig:2}a.
A seeding depth of $z_0 = -z(t_0) = \SI{4}{\metre}$ was chosen to measure the \R{downward vertical displacement} of fluid initially close to, but below, the surface of the ocean.
Based on Figure~\ref{fig:2}a, four regions of strong \dz\ are observed in the system.
Figure~S4 (Supporting Information) demonstrates that the \dz\ fields are qualitatively similar down to a depth of $\SI{20}{\metre}$.

High-\dz\ regions in the upper ocean remain nearly depth independent for all three models despite spanning various length scales (Table~S1, Supporting Information) and this vertical correlation further justifies a search for surface TZs to locate vertical transport conduits.
The instantaneous Eulerian metrics calculated on or just below the ocean surface at time $t_0$ are presented in Figures~\ref{fig:2}b-d.
Regions with the largest magnitudes of these metrics are concentrated near the vorticity filaments (as seen in Figure~\ref{fig:1}a) and form sharper structures compared to \dz\ in Figure~\ref{fig:2}a.
This is due to the Eulerian TZs highlighting the most dynamic regions on the surface at the instant $t_0$.
In contrast, the \dz\ field is computed over a finite time interval, which allows more of the fluid to collect towards and downwell along the transport pathways, ultimately resulting in broader regions of high \dz.
The dilation rate and FTLE fields computed on the surface for $T_L = T_d = \SI{6}{\hour}$ are presented in Figures~\ref{fig:2}e,f.
These Lagrangian fields computed over a time interval have broader features that are more representative of the \dz\ field.

The TZs for each metric are overlaid over the respective fields in Figures~\ref{fig:2}b-f.
A comparison between the TZs of a metric and the high-\dz\ regions provides an indication of the ability of the metric computed on the ocean surface to identify regions of strongest sub-surface vertical displacement.
The Eulerian TZs observed in Figures~\ref{fig:2}b-d are restricted to the filamentary regions, whereas the Lagrangian TZs in Figures~\ref{fig:2}e,f are more reflective of the high-\dz\ regions in Figure~\ref{fig:2}a.
This suggests that the Lagrangian TZs are better at identifying regions of stronger downward vertical displacement than the Eulerian ones.

\subsection{Locating Vertical Displacement Conduits in Three Submesoscale-resolving Models}
\label{subsec:results_2}
In order to evaluate if the benefit of using Lagrangian TZs over Eulerian is robust, we repeat the analysis over longer displacement times on different types of simulations that resolve physical processes at different scales.
As was done for the turbulence resolving model, the \dz\ field was computed for particles seeded at \SI{4}{\metre} for the process study and operational models.
The displacement times for the larger scale models were $T_d = \SI{12}{\hour}$, a time scale that is relevant for physiological responses of biological populations that often operate on diel cycles~\citep{dusenberry99}.
As was the case for the turbulence resolving model, the Lagrangian TZs were computed for time intervals matching the displacement times (i.e., $T_L = T_d$).

\begin{figure}[t]
    \centering
    \includegraphics[width=\linewidth]{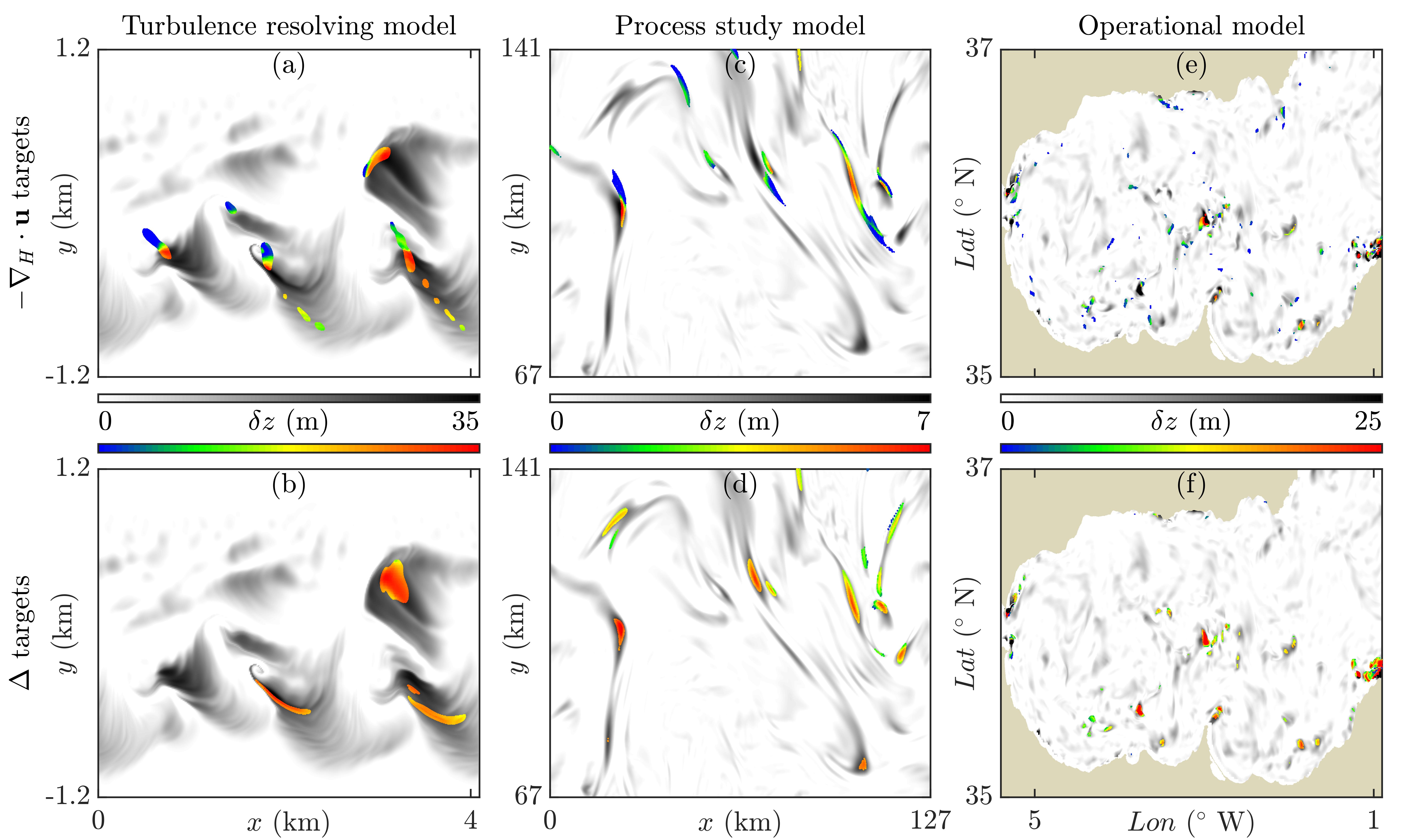}
    \caption{Target zones overlaid on \R{\dz\ field}s for the three models. The negative horizontal divergence $(-\nabla_H \cdot \mathbf{u})$ target zones (top) and dilation rate $(\Delta)$ target zones (bottom) overlaid (color) on \dz\ fields (grayscale) calculated for the turbulence resolving model (a) and (b), the process study model (c) and (d), and the operational model of the western Mediterranean (e) and (f). Target zone overlays are plotted using a colour map representing the \dz\ value underneath. Shades of blue represent regions where the target zones do not correspond to strong downward displacement and the red regions represent target zones that have identified \R{regions of strongest downward displacement}.}
    \label{fig:3}
\end{figure}

Figure~\ref{fig:3} presents the target zones obtained from the Eulerian divergence metric and Lagrangian dilation rate metric for each model overlaid on the respective \dz\ fields.
Dilation rate is the chosen Lagrangian metric since it is the best metric considered for identifying high \dz\ regions.
Divergence is chosen as the Eulerian metric as opposed to vertical velocity based on practical considerations, and the fact that dilation rate is the Lagrangian average of divergence along a particle trajectory.
The color of the TZ reflects the amount of underlying \R{downward vertical displacement} with red regions corresponding to where the TZ aligns with the high \R{\dz}\, and the blue regions to low \dz.
For the turbulence resolving model, the Eulerian TZs presented in Figure~\ref{fig:3}a fail to align with a significant portion of regions of high \dz.
The Eulerian TZs also correspond to regions with hardly any downward vertical displacement as seen by the dark blue regions in Figure~\ref{fig:3}a.
The Lagrangian TZs, as seen from the preponderance of the red shades in Figure~\ref{fig:3}b, better capture regions of high \dz\ for $T_L = \SI{6}{\hour}$ for the turbulence resolving model.
The difference in performance between the two TZs are more pronounced in the process study model (Figures~\ref{fig:3}c,d) for which $T_L = \SI{12}{\hour}$ is a longer time interval.
The Eulerian TZs rarely correspond to high-\dz\ regions, whereas the Lagrangian TZs capture all regions of strongest \R{downward vertical displacement}.
Qualitatively similar results are obtained for the operational model as well, with the divergence TZs failing to capture many regions of high \dz\ (Figure~\ref{fig:3}e), which are identified by the Lagrangian TZs (Figure~\ref{fig:3}f).
Similar overlays for the other three metrics are presented in Figure~S5 (Supporting Information), wherein we observe that the Lagrangian FTLE TZs, which are similar to the dilation rate results, also outperform the Eulerian TZs.

\begin{figure}[t]
    \centering
    \includegraphics[width=\linewidth]{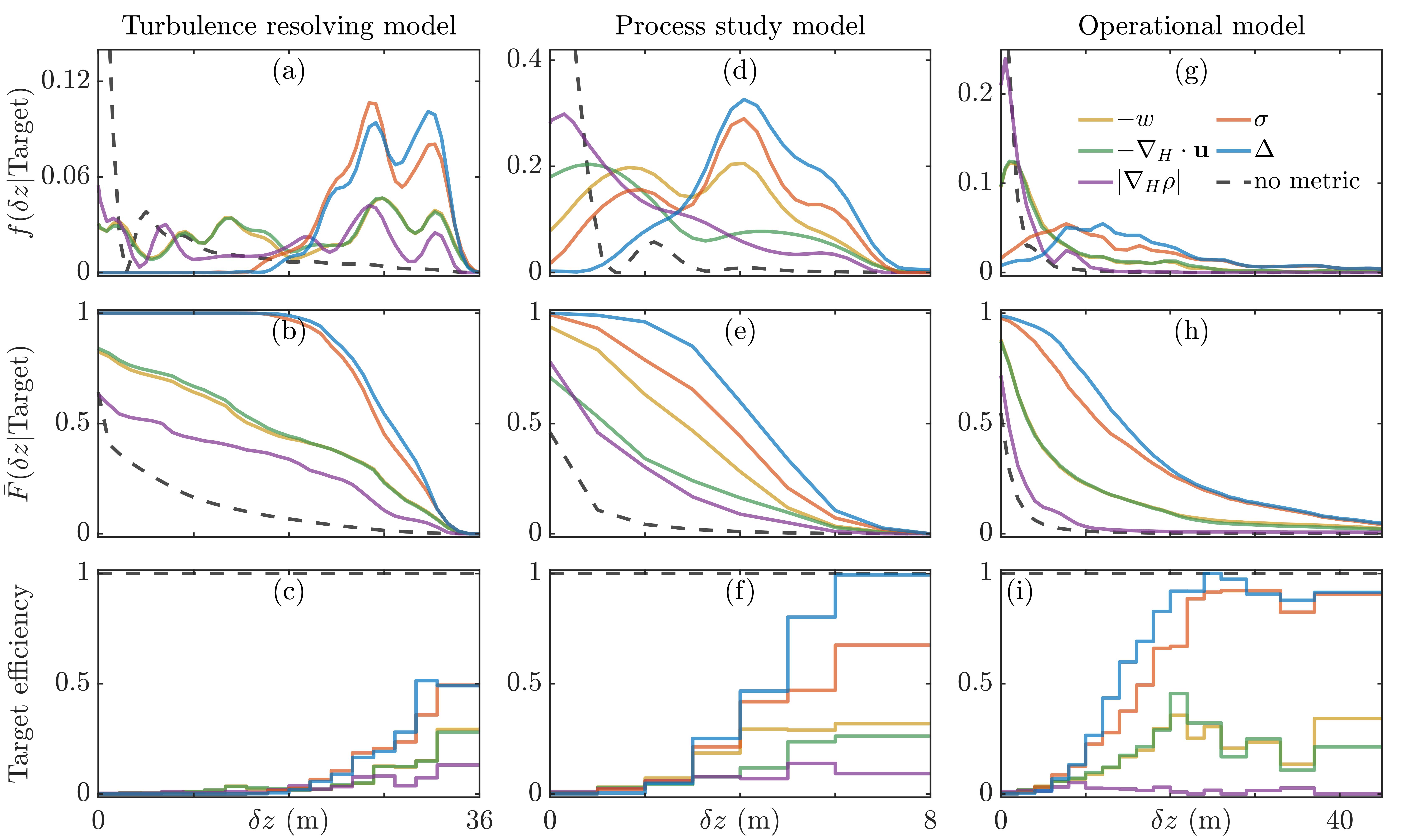}
    \caption{Skill metrics of the Eulerian and Lagrangian target zones.
    Probability density functions (PDFs) of \dz\ for each target zone $f(\delta z|$Target$)$ (top), corresponding complementary cumulative distribution functions $\bar{F}(\delta z|$Target$)$ (middle), and the target efficiency (bottom), plotted as a function of \dz, for the turbulence resolving model (left), process study model (centre) and operational model of the western Mediterranean (right).
    Legend for the metrics is presented in (g).
    Uniform bin-widths were chosen for the computation of the target efficiency, except for the bin corresponding to the largest \dz\ values, which has a lower bin-count.
    Skill curves for the no-metric case are also plotted as black dashed lines for reference, to highlight the benefit of using Eulerian and Lagrangian target zones as opposed to blindly sampling the entire domain.
    }
    \label{fig:4}
\end{figure}

To better quantify the degree of overlap between TZs and high-\dz\ regions in Figure~\ref{fig:3} and Figure~S5 (Supporting Information), skill metrics are computed and presented in Figure~\ref{fig:4}.
The probability distribution functions (PDFs) of \dz\ corresponding to the metric target zones $f(\delta z|$Target$)$ are presented in Figures.~\ref{fig:4}a,d,g.
This corresponds to the probability that a particle seeded below the surface TZ will be displaced downward by \dz.
Figures~\ref{fig:4}b,e,h present the corresponding complementary cumulative distribution functions (complementary CDFs) $\bar{F}(\delta z|$Target$)$, which indicate the probability of a particle seeded below the TZ displacing downward by at least \dz.
While evaluating the extent of \R{downward vertical displacement} corresponding to the TZs is helpful, we are also interested in the fraction of the strongest \dz\ zones that are identified by the TZs.
The ``target efficiency'' indicates the fraction of regions \R{corresponding to a downward vertical displacement of} \dz\ that are identified by a TZ, and is the third skill metric presented in Figures~\ref{fig:4}c,f,i.
The last two skill metrics, complementary CDF and target efficiency, can also be defined in terms of true positives ($TP$) or hits, false negatives ($FN$) or misses, and false positives ($FP$) or false alarms.
The complementary CDF is equivalent to $TP/(TP+FP)$ and the target efficiency is $TP/(TP+FN)$.
Figure~\ref{fig:4} also presents reference curves for the entire domain (no metric case), which corresponds to blindly sampling the domain, in black.
Note that the third skill metric, the target efficiency, in Figures~\ref{fig:4}c,f,i will always be 1 in this case since all regions with $\delta z>0$ are already included in the target zone (entire domain).

For the turbulence resolving model, the PDFs for the Lagrangian TZs in Figure~\ref{fig:4}a are skewed towards larger values of \dz, compared to a more uniform distribution for the Eulerian TZs.
This is in stark contrast to the PDF for the no metric case which is skewed towards \dz$=0$, indicating that on average there is no \R{downwelling} or upwelling in the domain.
The median \dz\ for the Lagrangian TZs $(\sim \SI{27}{\metre})$ is approximately double that of the Eulerian divergence and vertical velocity TZs $(\sim \SI{15}{\metre})$ and the density gradient TZs performed the worst with a median of only $\SI{6}{\metre}$.
The benefit of at least using an Eulerian metric is not to be dismissed, however, since the median \dz\ for the divergence TZs is \SI{14}{\metre} more than the value for the no metric case (less than $\SI{1}{\metre}$), which is of the same order of magnitude as the gain when using the Lagrangian dilation rate TZs compared to divergence TZs.
The Lagrangian TZs rarely identify regions of low \dz, as indicated by $f(\delta z|$Target$)=0$ for $\delta z<\SI{15}{\metre}$.
The PDFs for Eulerian TZs have non-zero values for $\delta z=0$, indicating that a small fraction of these target zones result in no \R{downwelling} or small amounts of upwelling.
This is not surprising since density fronts are sites of strong downwelling as well as upwelling~\citep{verma19}.
The complementary CDFs presented in Figure~\ref{fig:4}b demonstrate that the Lagrangian TZs have a consistently higher probability of predicting any amount of \R{downward vertical displacement} \dz, with the dilation rate marginally outperforming the FTLE.
For example, Lagrangian TZs identify $\delta z>\SI{20}{\metre}$ more than 90\% of the time, as opposed to 37\% for the Eulerian TZs, even though \dz\ of this magnitude occurs in only 5\% of the domain.
Furthermore, as observed in Figure~\ref{fig:4}c, Lagrangian TZs identify approximately 50\% of the strongest \dz\ regions, despite targeting only 1\% of the domain.
Across all three skill metrics, there is clear benefit of using the Lagrangian TZs over the Eulerian TZs.
Another interesting observation is that while the horizontal density gradient reveals the location of oceanic fronts, which are sites of enhanced vertical motion, the metric performs worse than the other Eulerian metrics at identifying regions of strongest \dz.

The benefit of using Lagrangian TZs is also observed for the larger scale models in Figures~\ref{fig:4}d-i.
For the $\SI{12}{\hour}$ analysis of the process study model, Figure~\ref{fig:4}d demonstrates that the Lagrangian dilation rate TZs are more than twice as likely to predict regions with $\delta z=\SI{4}{m}$ compared to the Eulerian divergence TZs, which is half the maximum value of \dz\ for the model.
The Lagrangian TZs are more likely to locate regions of large downward displacements, as indicated by Figure~\ref{fig:4}e, and the median \dz\ corresponding to Lagrangian TZs $(\sim \SI{4}{\metre})$ are four times the value of the Eulerian divergence TZs $(\sim \SI{1}{\metre})$.
Lagrangian TZs rarely identify regions of upwelling in the process study model with $\bar{F}(\delta z = 0|$Target$)\sim1$, as opposed to $\sim0.7$ for the Eulerian ones.
Finally, as presented in Figure~\ref{fig:4}f, the Lagrangian TZs identify majority (dilation rate TZs identify 99.4\%) of the \R{regions of strongest downward vertical displacement} where \dz$>\SI{6}{\metre}$ (which constitute only 0.1\% of the domain) while the Eulerian TZs continue to miss many of these regions, capturing less than $27\%$.
Amongst the Eulerian metrics, the vertical velocity TZs perform better than the other Eulerian metrics in this model.
This is because the vertical velocity was computed slightly below the ocean surface for numerical accuracy (discussed in the methods section).
The effect of the small vertical displacement in the depth where $-w$ TZs are computed is magnified as a result of the relatively small amounts of \dz\ in the process study model.
It is important to remember, however, that obtaining accurate vertical velocities in the real ocean is difficult, and hence this curve should only be used as a barometer for the other metrics.

The strongest demonstration of the benefit of using Lagrangian TZs is in the most realistic setting, the data-assimilative real-time operational model of the western Mediterranean.
For the $\SI{12}{\hour}$ analysis, the PDFs for Eulerian TZs (Figure~\ref{fig:4}g) have a large peak near $\delta z=0$, whereas the curves for Lagrangian TZs are flatter with a peak near $\SI{12}{\metre}$.
The median \dz\ detected by the Lagrangian TZs $(\sim \SI{12}{\metre})$ is four times larger than the Eulerian TZ results $(\sim \SI{3}{\metre})$.
As seen in Figure~\ref{fig:4}h, Lagrangian TZs are more likely to predict large downward vertical displacement.
For example, the Lagrangian TZs have a 30\% chance to locate \dz$\ge\SI{20}{\metre}$ as opposed to less than $10\%$ for the Eulerian TZs.
Not only are Lagrangian TZs more likely to locate large \dz, they also locate a majority of the regions of strongest \dz.
Specifically, dilation rate TZs locate $91\%$ of the regions where \dz$>\SI{40}{\metre}$ despite this only occurring in $0.1\%$ of the domain (Figure~\ref{fig:4}i).
\RR{In the case of the analysis of the operational model, the average mixed layer depth for the domain is approximately 10 m.
This value is in agreement with the average mixed layer depth of this region reported in literature, which lies between 5 and 30 m in April~\citep{d2005seasonal,mason2019new,tarry2021frontal}.
This indicates that some of the identified regions are possibly experiencing subduction in our analysis of the operational model of the western Mediterranean.}

\subsection{Effect of Varying the Seeding Depth}
\label{subsec:results_3}
In the previous sections, Lagrangian metrics computed on the surface of the ocean were shown to better identify signatures of strong sub-surface \dz\ compared to Eulerian metrics.
The three dimensional trajectories used to compute \dz\ for these analyses were initialized at a depth of $\SI{4}{\metre}$, and it is worthwhile investigating how the results change when particles are seeded further below the surface.
In this section, we investigate how the cumulative CDFs for the operational model change when the initialization depth is varied to understand how far below the surface the target zones are valid.

\begin{figure}[t]
    \centering
    \includegraphics[width=\linewidth]{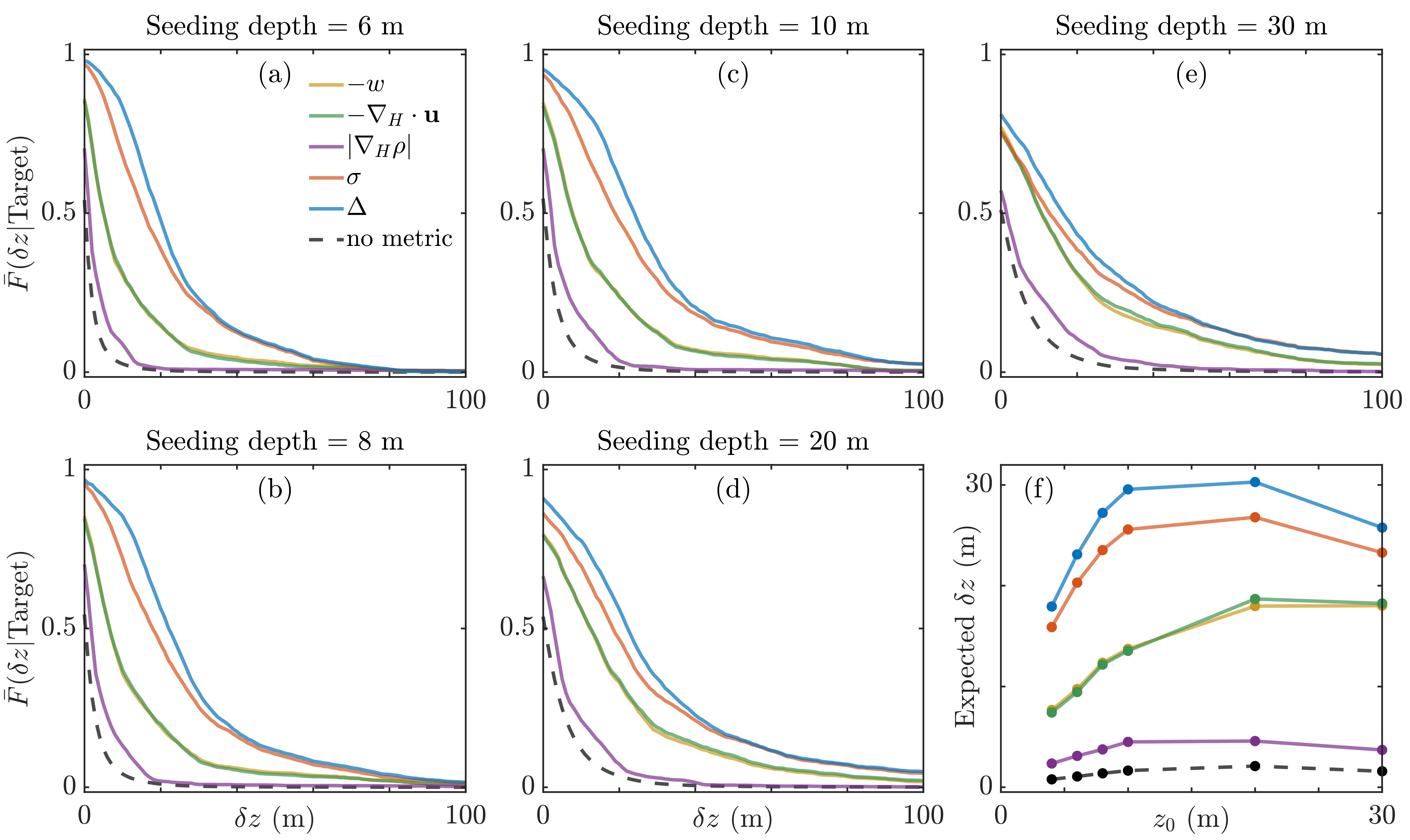}
    \caption{Variation of skill with seeding depth.
    (a-e) Complementary cumulative distribution functions $\bar{F}(\delta z|$Target$)$ for five different seeding depths at which particles are initizlized to compute \R{downward vertical displacement}, plotted as a function of the resulting \dz, for the operational model of the western Mediterranean.
    The target zones are computed using Lagrangian and Eulerian metrics computed on the ocean surface.
    (f) Expected \dz\ for TZs of each metric, plotted as a function of the seeding depth, $z_0$.
    The curve for the no metric case (black dashed) has also been plotted for reference.}
    \label{fig:5}
\end{figure}

Figure~\ref{fig:5} presents the variation of the cumulative CDF for the five metrics (as well as the reference curve for the no metric case) for five different seeding depths: $z_0=$\SIlist[list-units = single]{6;8;10;20;30}{\metre}.
In this figure, the \dz\ axis has been extended to \SI{100}{\metre} to accommodate the increased \R{downward vertical displacement} observed at deeper $z_0$.
The cumulative CDFs for $z_0\leq\SI{10}{\metre}$ are qualitatively similar, albeit with an increase in median \dz\ for the dilation rate TZs to $\SI{24}{\metre}$ for $z_0 = \SI{10}{\metre}$ and a small decay in $\bar{F}(\delta z = 0|$Target$)$.
The latter suggests a growing frequency of incorrectly targeting regions with small amounts of upwelling.
For deeper \R{\dz}\ measurements, $\bar{F}(\delta z = 0|$Target$)$ decays to 0.91 and 0.81 for $z_0 = $\SIlist{20;30}{\metre}, respectively, indicating a further increase in erroneously identifying upwelling regions as downwelling.
For dilation rate TZs, the median \dz\ for $z_0 = $\SIlist{20;30}{\metre} are also smaller at \SIlist{22;17}{\metre}, respectively, which means that the correlation between the surface signatures of the conduits, identified using the metrics, and their locations at deeper layers decay for $z_0>\SI{10}{\metre}$.
This is because the vertical displacement conduits located near density fronts are not strictly vertical (i.e., they are located at an angle with the vertical that is dependent on the front angle), and as a result the boundaries of the conduits below the surface may be horizontally shifted from their surface signatures.
For the two deepest $z_0$, the benefit of using Lagrangian metrics over Eulerian ones decreases, since the improvement is overshadowed by the decrease in correspondence between the surface and sub-surface.
The median \dz\ for the entire domain (i.e., no metric case) increases, however, which suggests that even though the magnitude of \R{vertical displacement} increases, a decay in the correspondence between surface metrics and the sub-surface dynamics results in reduced benefit of using the surface TZs.
While the benefit of using a Lagrangian target zone decays for deeper $z_0 = \SI{30}{\metre}$, using a surface target zones (with the exception of the density gradient TZs) is still more beneficial than randomly sampling the domain.
The variation with seeding depth $z_0$ in the skill associated with each metric is summarized using the expected \dz\ for each target zone, presented in Figure~\ref{fig:5}(f).
The expected \dz\ increases with seeding depth for all metrics until $z_0 = \SI{10}{\metre}$, since the particles originate close enough to the surface to have strong correspondence to surface signatures, while being deep enough to displace further downward, faster.
From $z_0 = \SI{10}{\metre}$ to \SI{20}{\metre}, the expected \dz\ remains nearly constant for the Lagrangian metrics $\sigma$ and $\Delta$, but not for $-w$ and $-\nabla_H \cdot \mathbf{u}$.
This occurs despite the TZs locating conduits that displace water parcels deeper because the extent of non-downwelling regions located by the TZs increases with $z_0$ (as seen by the decrease in $\bar{F}(\delta z = 0|$Target$)$) for $\sigma$ and $\Delta$, but remains nearly the same for $-w$ and $-\nabla_H \cdot \mathbf{u}$.
Regardless, the magnitudes of expected \dz\ for the Lagrangian TZs remain 50\% more than the values for the two Eulerian TZs.
The expected \dz\ does not improve any further for $z_0 = \SI{30}{\metre}$, even for $-w$ and $-\nabla_H \cdot \mathbf{u}$ TZs, because the correspondence between the surface and the sub-surface breaks down.

To better understand the the weakening relationship between surface metrics and \dz\ for particles starting deeper below the surface, we evaluate the correlation between two-dimensional trajectories that are artificially restricted to the ocean surface, like drifter data would be, and the three-dimensional trajectories of particles seeded at various depths.
To evaluate this, we compute the horizontal separation, $d_h$, between the two-dimensional surface trajectories and the surface projections of the fully three-dimensional trajectories at the final instant of the $\SI{12}{\hour}$ advection interval as a function of the particles being inside or outside the dilation rate target zone.

\begin{figure}[t]
    \centering
    \includegraphics[width=\linewidth]{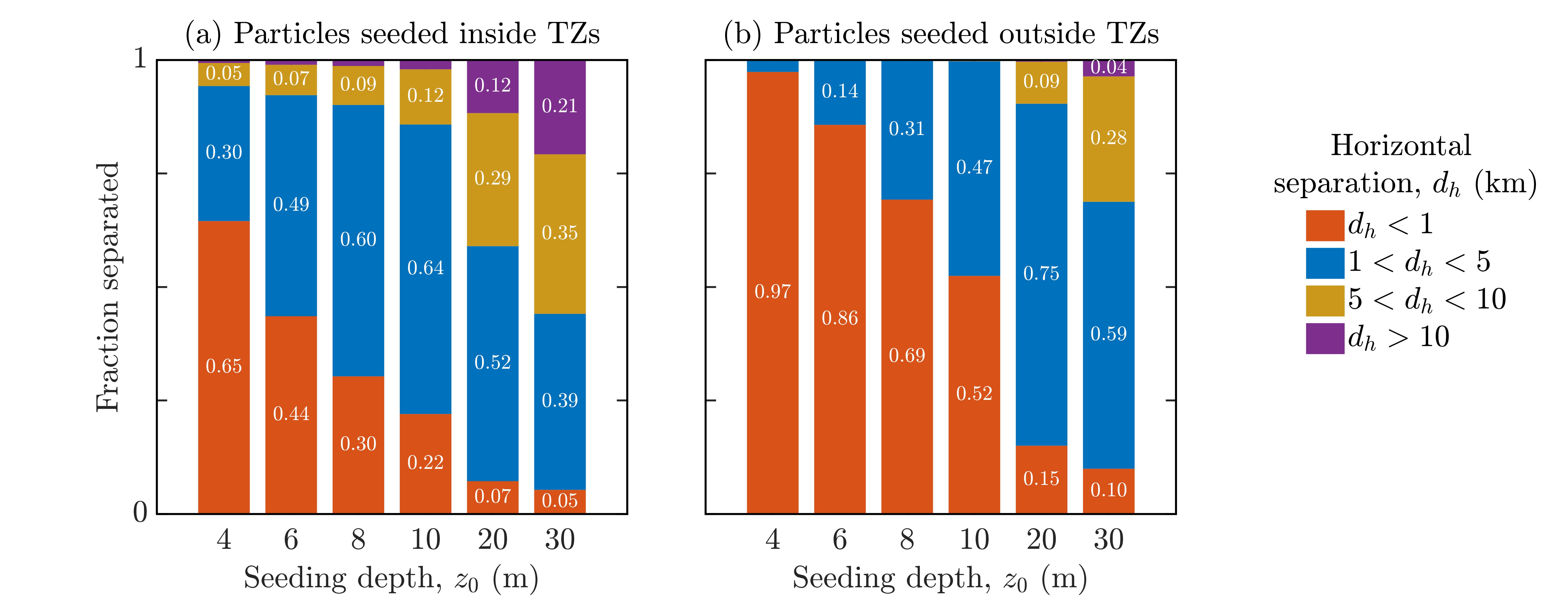}
    \caption{Statistics of horizontal separation ($d_h$) between two-dimensional surface trajectories and the three-dimensional trajectories of particles seeded (a) inside and (b) outside the dilation rate target zones.
    The three dimensional trajectories are computed for particles seeded at six different depths: $z_0 = $\SIlist{4;6;8;10;20;30}{\metre}. The fraction of trajectories with $d_h<\SI{1}{\kilo\metre}$ (red bars), $\SI{1}{\kilo\metre}<d_h<\SI{5}{\kilo\metre}$ (blue bars), $\SI{5}{\kilo\metre}<d_h<\SI{10}{\kilo\metre}$ (yellow bars) and $d_h>\SI{10}{\kilo\metre}$ (purple bars) are presented for each $z_0$.}
    \label{fig:6}
\end{figure}

Figure.~\ref{fig:6} compares the statistics of $d_h$ for particles seeded at the six different depths. The fraction of particles seeded inside (Figure~\ref{fig:6}a) and outside (Figure~\ref{fig:6}b) the dilation rate target zones that have horizontal separations $d_h<\SI{1}{\kilo\metre}$, $\SI{1}{\kilo\metre}<d_h<\SI{5}{\kilo\metre}$, $\SI{5}{\kilo\metre}<d_h<\SI{10}{\kilo\metre}$ and $d_h>\SI{10}{\kilo\metre}$ are presented for each $z_0$.
For particles released nearest to the ocean surface, the majority of three dimensional trajectories remain less than $\SI{1}{\kilo\meter}$ away from the corresponding surface trajectory.  There is a greater separation for particles that are released inside the TZs as $5\%$ of trajectories have a final $d_h>\SI{5}{\kilo\meter}$.  As particles are seeded deeper in the domain the similarity between the surface and three dimensional trajectory monotonically decreases.  Despite this, the trend that trajectories outside the TZs are approximately twice as likely to remain less than $\SI{1}{\kilo\meter}$ is consistent.  For seeding depths up to $\SI{10}{\meter}$, a negligible fraction of trajectories released outside the TZs have greater than $\SI{5}{\kilo\meter}$ separation, which confirms that the vast majority of the upper ocean moves predominately in a two dimensional manner with the surface trajectories being well correlated with the horizontal projections of the fully three dimensional trajectories.
Meanwhile greater than $\SI{5}{\kilo\meter}$ separation is observed between $5$ and $14\%$ of the time within the dilation rate TZs, which correspond to sites of increased \R{vertical displacement} as discussed in the previous sections.

Even within the TZs, larger horizontal separation corresponds to greater \dz.
For particles seeded inside TZs at $z_0 = \SI{4}{\metre}$, the median \dz\ corresponding to $d_h<\SI{1}{\kilo\metre}$ is only $\SI{12}{\metre}$, but it increases to $\SI{23}{\metre}$ for $\SI{1}{\kilo\metre}<d_h<\SI{5}{\kilo\metre}$ and $\SI{25}{\metre}$ for $\SI{5}{\kilo\metre}<d_h<\SI{10}{\kilo\metre}$.
The median \dz\ for the trajectories that separate the most ($d_h>\SI{10}{\kilo\metre}$, which corresponds to only 0.7\% of the target zones), however, is $\SI{53}{\metre}$, confirming that the trajectories that are displaced downward the most correspond to the largest horizontal separation.
Vertical displacement along the coherent conduits that have a preferred orientation likely plays a role here, since enhanced vertical velocity alone need not correspond to increased horizontal separation $d_h$.
For a seeding depth of $z_0 = \SI{10}{\metre}$, 2\% of particles seeded inside the TZs result in $d_h>\SI{10}{\kilo\metre}$, whereas only 0.3\% of those seeded outside separate even $\SI{5}{\kilo\metre}$.
For the larger seeding depths of \SI{20}{\metre} and \SI{30}{\metre}, a non-trivial fraction of trajectories initialized outside the dilation rate TZs obtained using surface metrics separate more that \SI{5}{\kilo\metre}.
These trajectories that separate the most are still linked to particles that are vertically displaced more than the average, but the horizontal shift in the location of the conduits at this depth compared to the surface results in the surface TZs not capturing some of the trajectories that are the displaced the most vertically.

\subsection{Effect of Varying the Time Intervals of Analysis.}
\label{subsec:results_4}
\begin{figure}[t]
    \centering
    \includegraphics[width=\linewidth]{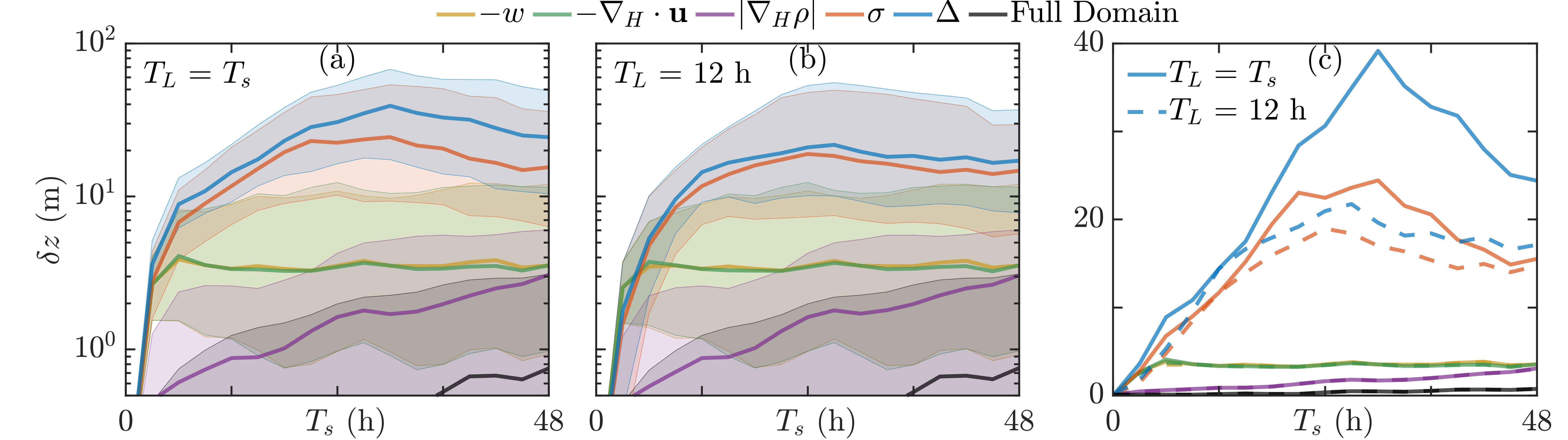}
    \caption{Effect of analysis time intervals on median \dz\ for the five target zones, applied to the operational model of the western Mediterranean. Median \dz\ and interquartile ranges are plotted on a logarithmic-scale as a function of displacement time ($T_d$), with the time interval for Lagrangian target zone calculation ($T_L$) set to (a) the same value as $T_d$ and (b) a value of \SI{12}{\hour}. (c) Median \dz\ curves (linear-scale) for target zones corresponding to the two Lagrangian metrics FTLE and dilation rate from (a) and (b) for comparison. The median \dz\ for the entire domain (black) without any targeting remains close to zero over the $\SI{48}{\hour}$ interval, highlighting the importance of using efficient targeting methods.}
    \label{fig:7}
\end{figure}

We have shown for all three models that the Lagrangian TZs identify regions of high \dz\ better than Eulerian TZs for a fixed time interval.
The length of the time interval of analysis plays a significant role in both the \R{\dz}\ and Lagrangian metric calculations, so it is worthwhile considering how the TZs perform for varying integration times and in cases where the Lagrangian metrics are calculated over a different time interval than the \R{\dz}.
For practical purposes, factors such as the frontal strength and eddy dynamics may affect the length of time over which subduction events occur.
Hence, we consider a range of displacement times $T_d$ applied to the operational model of the western Mediterranean, for particles seeded at a depth of \SI{4}{\metre}.
Figure~\ref{fig:7}a presents the effects of varying $T_d$, while maintaining $T_L=T_d$, on the median \dz\ and the interquartile range for each of the five TZs.
Note that varying $T_L$ does not affect the Eulerian TZs since these metrics are computed only at the initial instant.
The \dz\ field, however, does vary with a change in $T_d$, and hence the median associated with Eulerian TZs will depend on $T_d$.
As the length of the time interval increases, the median \dz\ for the Lagrangian dilation rate TZs rapidly increases before peaking at $\sim\SI{40}{m}$ at $T_L=\SI{30}{h}$.
The peak may indicate that there is a slowing of vertical displacement after $T_d=T_L\sim\SI{40}{\hour}$, or that there is a limit to the correspondence of the surface signatures to the dynamics well below the surface.
The median \dz\ from the Eulerian TZs also increases but is never more than \SI{4}{m}, with the density gradient TZs yielding a lower median \dz\ than the other two metrics.
The difference between Lagrangian and Eulerian TZs is so significant that almost the entire interquartile range for the Lagrangian dilation rate TZs is always larger than the third quartile for the Eulerian TZs.
For reference, the median \dz\ and the interquartile range have also been plotted in black for the case without any target zone (i.e., sampling the entire domain).
The median \dz\ in this case is increasing but remains close to zero, indicating that despite extending the time interval considered there remains on average negligible net upward or downward motion associated with particles that start at a depth of \SI{4}{\metre}.

From a practical perspective, experiments may need to be performed over displacement times longer than the time interval over which accurate model flow fields are available, since forecasts are less reliable further into the future.
Hence, the efficacy of Lagrangian target zones computed over the time-scale of half a day for various displacement times is investigated.
Figure~\ref{fig:7}b highlights that, for a constant $T_L = \SI{12}{\hour}$, the Lagrangian TZs still achieve larger median \dz\ and interquartile ranges than Eulerian ones for $T_d>\SI{6}{\hour}$.
Comparing the median \dz\ curves for the Lagrangian TZs when either $T_L=\SI{12}{h}$ or $T_L=T_d$, Figure~\ref{fig:7}c demonstrates that although it is better to compute the Lagrangian TZs over the same time interval as \R{\dz}\, a half-day computation yields median \dz\ values of the same order of magnitude.
When $T_L$ is held constant, although there is a local maximum in median \dz\ for the Lagrangian TZs around $T_d = \SI{24}{\hour}$ the median \R{\dz}\ plateaus for longer $T_d$.
For both studies, the Lagrangian TZs based on dilation rate consistently outperforms FTLE in identifying regions of high \dz.

\subsection{Ensemble Analysis of the Operational Model.}
\label{subsec:results_5}
Another challenge for using Lagrangian target zones is that the true velocity field is not known beforehand and there are uncertainties associated with the ocean model fields as a result of parametrisations of unresolved processes and specifications of initial and boundary conditions~\citep{lermusiaux06}.
To investigate the effectiveness of the TZs without knowledge of the actual velocity field, we perform an ensemble analysis of the western Mediterranean.
Specifically, 191 realizations of real-time forecasts of the western Mediterranean were first generated using the Error Subspace Statistical Estimation schemes~\citep{lermusiaux02,mahadevan20}.
One realization was then randomly chosen to be the ``true'' outcome, which was the realization presented in the previous sections, while the remaining 190 were used to compute the ensemble TZs.
Similar results are obtained when selecting other realizations to be the ``truth''.
The ensemble TZs for each metric are the regions with values in the ninety-ninth percentile of the ensemble averaged field, obtained by averaging the metric over the 190 realizations.
The ensemble TZs were then compared to the high-\dz\ regions in the true realization to evaluate the efficacy of each metric to be used in a predictive sense.
For clarity, only the results for Eulerian divergence and the Lagrangian dilation rate are presented in this study.
Results for the other metrics are presented in Figure~S6 (Supporting Information).

\begin{figure}[t]
    \centering
    \includegraphics[width=\linewidth]{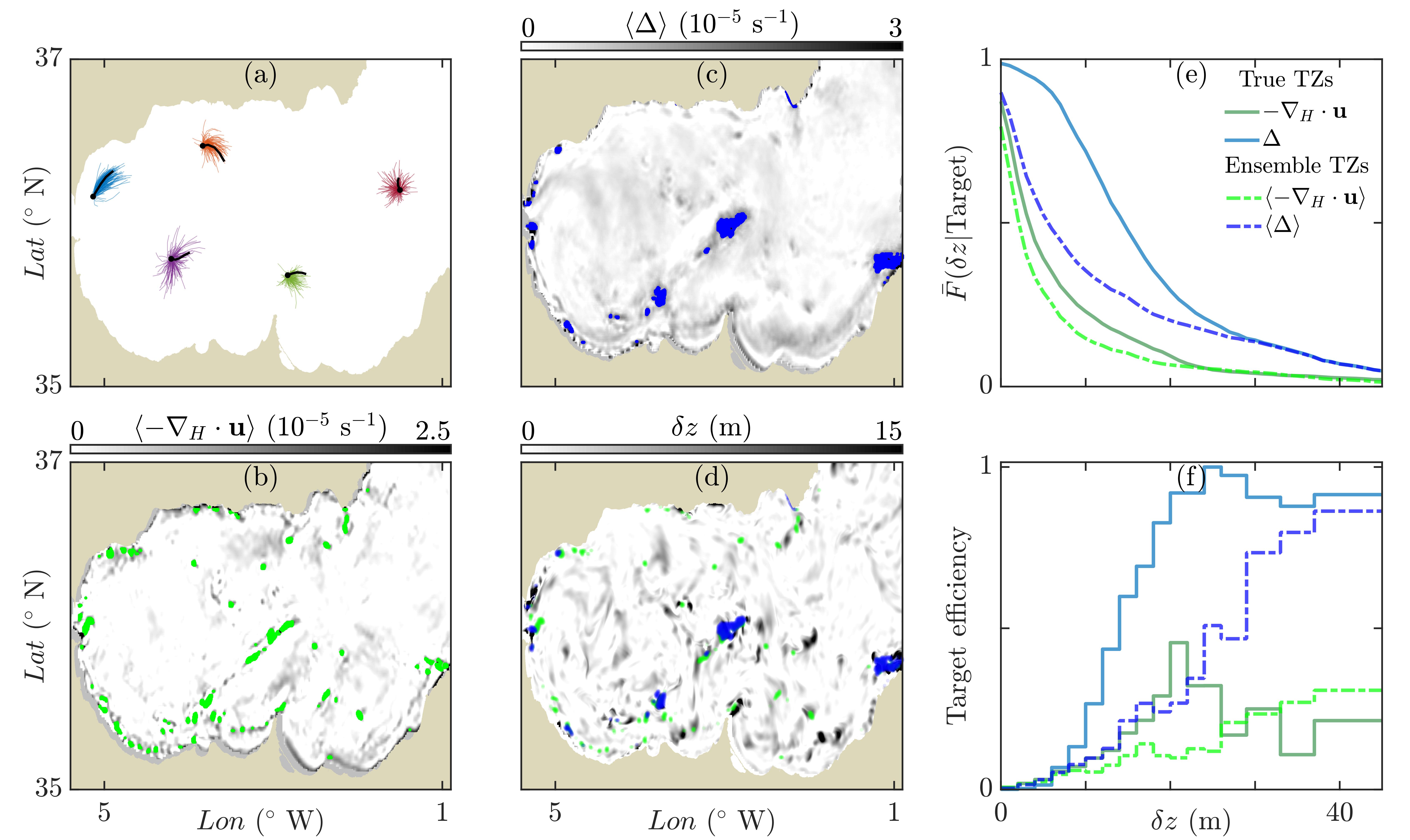}
    \caption{Ensemble analysis of the operational model of the western Mediterranean. (a) Surface trajectories of particles released at five locations for 190 realizations (colour); black lines correspond to trajectories from the ``true'' realization. Ensemble averaged (b) negative horizontal divergence, $-\nabla_H \cdot \mathbf{u}$, and (c) dilation rate, $\Delta$, fields, with corresponding target zones overlaid. (d) Ensemble target zones obtained in (b) and (c) overlaid on the true \R{\dz\ field}. (e) Complementary CDF of \dz\ for the two ensemble-averaged (dashed lines) and the true (solid lines) target zones. (f) Target efficiency of high-\dz\ regions in the true realization that are identified by the two ensemble (dashed lines) and the true (solid lines) target zones.}
    \label{fig:8}
\end{figure}

Figure~\ref{fig:8}a presents surface trajectories of particles seeded at five different locations for each of the 190 realizations, as well as the true realization.
The high degree of variability in surface trajectories, as evidenced by relative dispersion of up to $\SI{40}{\kilo\metre}$, reflects the variability in the ensemble velocity fields.
The uncertainties were forecast to be large due to the limited observations available in real-time and the significant submesoscale variability.
The accuracy of these real-time ensemble forecasts was validated using real drifter trajectories and other data sets (\url{http://mseas.mit.edu/sea_exercises/CALYPSO/2019/}).
The divergence and dilation rate fields were computed for each realization over half a day (for this analysis, $T_d = T_L = \SI{12}{\hour}$), and the ensemble average of these fields is presented in Figures~\ref{fig:8}b,c.
The TZs obtained from the ensemble-averaged fields are overlaid in black.
The variability in the Eulerian metric has averaged out to give negligible values in the bulk of the Mediterranean in Figure~\ref{fig:8}b, which results in the near-absence of Eulerian TZs away from the shore.
For reference, the peak divergence value for the true realization is \SI{0.33e-3}{\per\second}, and the ensemble-average at that position is \SI{0.2e-4}{\per\second}.
The resulting TZs in Figure~\ref{fig:8}b are significantly different from the green regions for the TZs using the ``true'' velocity presented in Figure~\ref{fig:3}e.
The Lagrangian TZs are more robust to model uncertainties as the ensemble-averaged field in Figure~\ref{fig:8}c yields well-defined TZs in the bulk of the Mediterranean that are reminiscent of the blue TZs in Figure~\ref{fig:3}f.
To elucidate, the overlap between ensemble TZs and true TZs for the Lagrangian metric dilation rate is better with a $34\%$ overlap, compared to a $16\%$ overlap for the Eulerian metric divergence.

Figure~\ref{fig:8}d presents the TZs from the ensemble-averaged fields overlaid on the \dz\ field from the true realization.
The dilation rate TZs align better with the high \R{\dz}\ regions compared to the divergence TZs, demonstrating that the Lagrangian dilation rate captures regions of high \R{downward vertical displacement} better than its Eulerian counterpart, the divergence.
This result is further demonstrated by $\bar{F}(\delta z|$Target$)$ in Figure~\ref{fig:8}e.
Although the ensemble Lagrangian TZs perform better than the ensemble Eulerian TZs, a comparison with the true Lagrangian TZs reveals the cost of not having the true velocity field.
Understandably, the ensemble Lagrangian TZs performs worse than the true Lagrangian TZs for $\delta z < \SI{35}{\metre}$.
What is remarkable, however, is that the ensemble-averaged Lagrangian TZs are almost as good as the true TZs for larger values of \dz.
This indicates that signatures of the strongest downwelling regions were captured by the Lagrangian TZs in most of the 190 realizations, as can be seen in Figure~\ref{fig:8}f.
The ensemble-averaged Lagrangian TZs are able to identify $\sim50\%$ of the regions of strongest \dz, similar to the true Lagrangian TZs.
Our result that target zones corresponding to the Lagrangian metric dilation rate is capable of detecting persistent regions of high \dz\ is complementary to the result that Lagrangian coherent structure calculations using the Lagrangian metric FTLE is robust to model error~\citep{haller02,lermusiaux2005dynamics,lermusiaux06}.
The ensemble-averaged and true Eulerian TZs are comparable for $\delta z>\SI{30}{\metre}$, but fail to perform as well as the ensemble-averaged Lagrangian TZs for all values of \dz.
This means that the Lagrangian TZs computed without knowledge of the true fields can identify the true high-\dz\ regions better than the Eulerian TZs computed using the true velocity fields.
Therefore, if a realistic ensemble forecast is available, Lagrangian TZs based on this forecast can be used as predictors of regions of strong sub-surface \R{downward vertical displacement} more accurately than Eulerian TZs using the current velocity field.

\section{Discussion}
Vertical transport plays a crucial role in the dynamics of the upper ocean, and the use of Lagrangian surface target zones enhances the ability to locate conduits of downward displacement.
We proposed the use of target zones (TZs) based on two Lagrangian metrics to identify regions of strong \R{downward vertical displacement} (\dz), and compared them with three Eulerian metric TZs for a turbulence resolving model of a density front, a process study model of a density front, and a data-assimilative real-time operational model of the western Mediterranean.
The five metrics were computed on the ocean surface with only surface fields, and the regions with values in the ninety-ninth percentile were chosen to be TZs for potential high-\dz\ conduits.
The performance of each TZ was evaluated by comparing their location with the \dz\ field computed below the ocean surface.
Although all five metrics highlighted regions of \R{downward vertical displacement} better than randomly sampling the domain, the Lagrangian TZs overlap better with regions of high \dz.
This resulted in Lagrangian TZs having a higher probability of predicting regions of high \dz, regardless of the length scales in the model, and larger median \R{\dz}.
Lagrangian TZs outperformed Eulerian ones for the full range of time intervals analysed, with median \dz\ up to an order of magnitude larger for the operational model.
Not only are the Lagrangian TZs more likely to identify regions of high \dz, but they also identify a large fraction of the strongest \dz\ regions.
Almost half of the \R{regions with strongest downward vertical displacement} in the operational model align with the Lagrangian TZs, despite the target zones constituting only $1\%$ of the entire domain.

Having demonstrated the consistent benefit of using Lagrangian TZs to identify regions of strong sub-surface \R{vertical displacement}, we considered two practical concerns of employing this method: (a) availability of forecast fields for displacement times of interest, and (b) sensitivity of results to model uncertainty.
In practice, experiments might need to be performed over displacement times longer than the time interval for which accurate forecast fields are available or target zone computation is reliable.
Analysis of the operational model demonstrated that a 12-hour Lagrangian analysis is sufficient to predict high-\dz\ regions for up to two days, with longer analysis only resulting in modest improvements.
Since these methods may be used in a predictive sense, the actual velocity field may not be available for the Lagrangian analysis.
The ensemble analysis of the western Mediterranean demonstrated that even when the true velocities are not known, and only an ensemble forecast is available, Lagrangian TZs are able to predict high-\dz\ regions.
Specifically, we demonstrated that Lagrangian TZs computed from an ensemble forecast fields perform nearly as well as the true Lagrangian TZs for regions with $\delta z>\SI{35}{m}$.
Interestingly, the ensemble Lagrangian TZs performed better than the true Eulerian TZs, indicating that a Lagrangian analysis of realistic forecasts is better at identifying the regions of strongest \dz\ than Eulerian targeting based on the actual velocity fields.

The ability of Lagrangian metrics to predict high-\dz\ regions is indicative of their link to vertical transport in the ocean, and opens up the possibility for more detailed studies of the phenomenon from a Lagrangian perspective.
\R{Our analysis has located strong vertical displacement  exclusively using surface information.
To identify where particles originally in the mixed layer are transported below it, accounting for the mixed layer depth in conjunction with the target zones may provide a means for differentiating between vertical displacement and subduction.}
There are potential benefits of considering combinations of the Lagrangian metrics and measurable ocean properties.
For instance, the Lagrangian dilation rate TZs could be used in conjunction with the measured density gradient indicative of an oceanic density front to only target the vertical displacement regions associated with frontal submesoscale processes, in order to identify the coherent subduction conduits.
A drifter release strategy leveraging the forecasted TZs and mixed layer depths could then be devised to more confidently identify surface signatures of subduction.
Surface drifter trajectories, however, can differ based on the type of drifter due to buoyancy and windage effects~\citep{olascoaga20}, and it remains to be determined how drifter convergence zones~\citep{molinari75,berta16} reflect the ocean surface TZs.
Investigation of how TZs are modified by inertial effects when metrics like divergence and dilation rate are estimated using drifters will allow us to discard spurious surface signatures of convergence.
These developments could potentially allow for direct identification of sites of strong vertical displacement and subduction based on forecasts and drifter based measurements, resulting in significant savings and a more efficient investigation of the ocean physics.

\section*{CRediT authorship contribution statement}
\textbf{H. M. Aravind:} Data curation, Formal analysis, Investigation, Methodology, Software, Validation, Visualization, Writing --- original draft, Writing --- review and editing.
\textbf{Vicky Verma:} Investigation, Writing --- review and editing.
\textbf{Sutanu Sarkar:} Funding acquisition, Investigation, Resources, Writing --- review and editing.
\textbf{Mara A. Freilich:} Investigation, Writing --- review and editing.
\textbf{Amala Mahadevan:} Funding acquisition, Investigation, Resources, Writing --- review and editing.
\textbf{Patrick J. Haley:} Investigation, Writing --- review and editing.
\textbf{Pierre F. J. Lermusiaux:} Funding acquisition, Investigation, Resources, Writing --- review and editing.
\textbf{Michael R. Allshouse:} Conceptualization, Funding acquisition, Investigation, Methodology, Project administration, Resources, Supervision, Writing --- original draft, Writing --- review and editing.

\section*{Declaration of competing interest}
The authors declare that they have no known competing financial interests or personal relationships that could have appeared to influence the work reported in this paper.

\section*{Acknowledgements}

The datasets generated and/or analysed can be accessed at the following DOI: 10.5281/zenodo.5142855. The authors would like to acknowledge Hieu T. Pham for his assistance in developing the turbulence resolving model.  H.M.A. and M.R.A. were supported by the ONR Grant N00014-18-1-2790. V.V. and S.S. were supported by ONR grant N00014-18-1-2137. M.A.F. and A.M. were supported by the ONR grant N00014-16-1-3130. P.F.J.L. and P.J.H. were supported by the ONR grant N00014-18-1-2781.



\bibliographystyle{elsarticle-harv}
\bibliography{main_final}

\end{document}


\begin{frontmatter}

\title{Lagrangian surface signatures reveal upper-ocean vertical displacement conduits near oceanic density fronts - Supporting Information}

\author[1]{H. M. Aravind}
\author[2]{Vicky Verma}
\author[2]{Sutanu Sarkar}
\author[3]{Mara A. Freilich}
\author[4]{Amala Mahadevan}
\author[5]{Patrick J. Haley}
\author[5]{Pierre F. J. Lermusiaux}
\author[1]{Michael R. Allshouse\corref{cor1}}
\cortext[cor1]{m.allshouse@northeastern.edu}

\affiliation[1]{organization={Mechanical and Industrial Engineering, Northeastern University},
            addressline={}, 
            city={Boston},
            postcode={02115}, 
            state={MA},
            country={USA}}
\affiliation[2]{organization={Mechanical and Aerospace Engineering, University of California San Diego},
            addressline={}, 
            city={La Jolla},
            postcode={92093}, 
            state={CA},
            country={USA}}
\affiliation[3]{organization={MIT–WHOI Joint Program in Oceanography},
            addressline={}, 
            city={Cambridge},
            postcode={02139}, 
            state={MA},
            country={USA}}
\affiliation[4]{organization={Woods Hole Oceanographic Institution},
            addressline={}, 
            city={Woods Hole},
            postcode={02139}, 
            state={MA},
            country={USA}}
\affiliation[5]{organization={Department of Mechanical Engineering, Massachusetts Institute of Technology},
            addressline={}, 
            city={Cambridge},
            postcode={02139}, 
            state={MA},
            country={USA}}

\end{frontmatter}

\noindent\textbf{Introduction}

This file includes six text-blocks, six figures and a table that is paired with Figure~S4. Each block of text contains a brief discussion on the corresponding figure: 
\begin{enumerate}
    \item Instantaneous limit of Lagrangian metric FTLE, the iLE
    \item Effect of varying the size of the target regions
    \item Vertical displacement, surface metrics and target zones for the fully turbulent flow
    \item Variation of \dz\ field with depth for the turbulence resolving model
    \item Target zone overlays on \dz\ fields
    \item Ensemble analysis of the operational model
\end{enumerate}

\newpage
\noindent\textbf{Text S1. Instantaneous limit of Lagrangian metric FTLE, the iLE}

\begin{figure}[h]
    \centering
    \includegraphics[width=\linewidth]{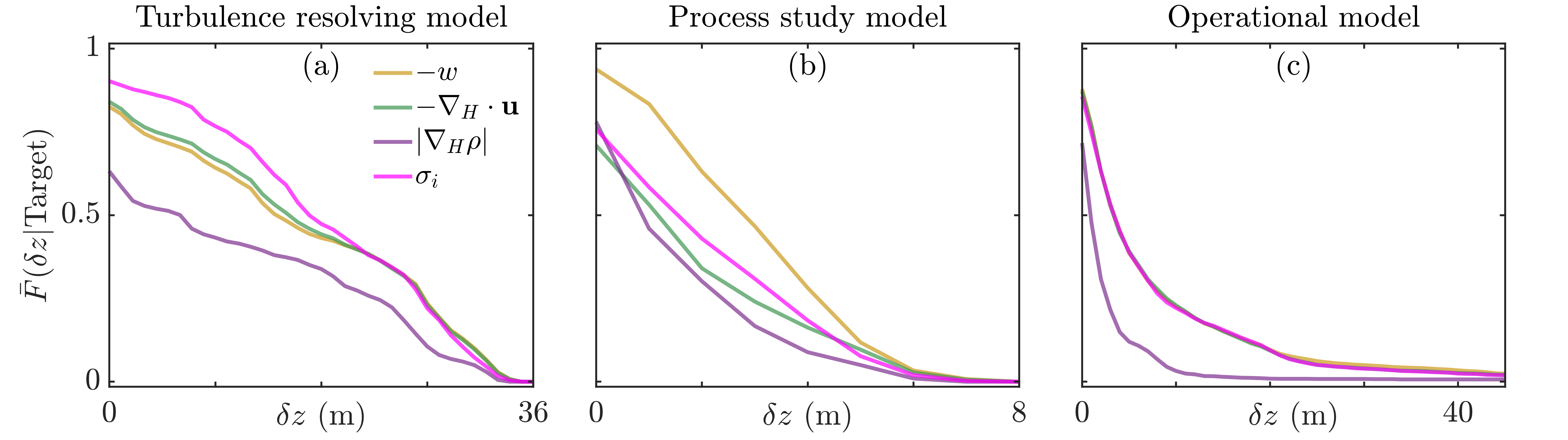}
    \caption[Instantaneous limit of Lagrangian metric FTLE, the iLE]{Complementary CDF of \dz\ for target zones derived from the instantaneous Lyapunov exponent ($\sigma_i$), the infinitesimal integration time limit of the finite-time Lyapunov exponent ($\sigma$), plotted for the three models. Curves for other Eulerian metrics presented in Fig.~4 of the main text are plotted for comparison}
    \label{fig:S2}
\end{figure}

The main study considers three Eulerian and two Lagrangian metrics for deriving surface target zones of sub-surface vertical displacement. The conclusion that Lagrangian metrics perform better than Eulerian ones motivates an investigation of the efficacy of the former at the infinitesimal integration time limit.
Figure~5 in the main text sheds some light into this, wherein we observe that the Lagrangian targets perform only as well as $-w$ and $-\nabla_H \cdot \mathbf{u}$ targets for very small $T_L$.  The infinitesimal integration time limit of the dilation rate is the divergence.
Here, we consider the theoretical infinitesimal integration time limit of the finite-time Lyapunov exponent, the instantaneous Lyapunov exponent~\cite{nolan20}. 
For this study, we define the metric as $\sigma_i = -s_1$, where $s_1$ is the smallest eigenvalue of the Eulerian rate of strain tensor.

Supplementary Fig.~\ref{fig:S2} presents the complementary CDF of \dz\ for target zones derived from $\sigma_i$, in addition to the ones for the other Eulerian metrics. 
For the two smaller-scale models, $\sigma_i$ performs slightly better than the $-\nabla_H \cdot \mathbf{u}$ TZs for \dz\ less than the median \dz. 
However, for larger \dz\, $\sigma_i$ targets are only as good as the $-w$ and $-\nabla_H \cdot \mathbf{u}$ target zones for all three models. 
The fact that the instantaneous limit of $\sigma$ performs no better than other Eulerian metrics reinforces our claim that performing an analysis over a finite time interval is essential to capture \dz\ that occurs over a period of time.

\newpage
\noindent\textbf{Text S2. Effect of varying the size of the target regions}

\begin{figure}[h]
    \centering
    \includegraphics[width=\linewidth]{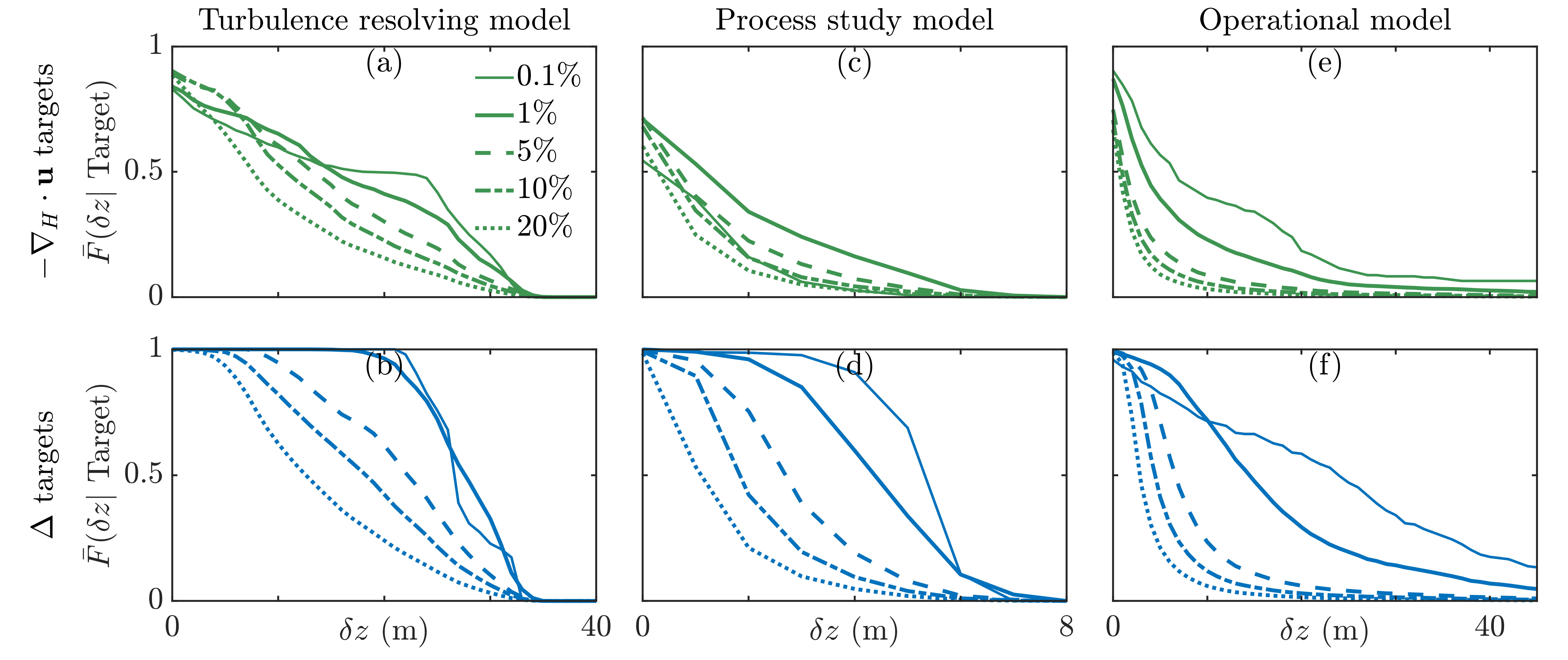}
    \caption[Effect of varying the size of the target regions.]{Effect of varying the size of the target regions. Variation of the complementary CDFs for the Eulerian target zone $-\nabla_H \cdot \mathbf{u}$ (top) and the Lagrangian target zone $\Delta$ (bottom) with size of respective target regions, expressed as a percent of the size of the domain. Results are presented for the turbulence resolving model (left), process study model (centre) and operational model of the western Mediterranean (right).}
    \label{fig:S3}
\end{figure}

The target zones for our study were chosen to be regions with the top 1\% values of the Lagrangian and Eulerian metric fields computed at the surface.
For an ocean model that spans a large area, however, the 1\% threshold might be too prohibitive due to the sparseness of the targets.
It might be the case that only the strongest \dz\ zones are highlighted even though other regions with significant vertical displacement exist.
Hence, it is worth investigating the effect of varying the 1\% threshold, and the results are presented for the Eulerian metric $-\nabla_H \cdot \mathbf{u}$ and the Lagrangian metric $\Delta$ for all three models analysed.  Similar results are obtained for the other metrics.

The complementary CDF of \dz\ for the horizontal divergence and dilation target zones for various target thresholds are presented in Supplementary Fig.~\ref{fig:S3}.
It is observed that as we increase the target threshold from the top 20\% to 1\%, targets become more likely to identify the stronger vertical displacement regions, resulting in increasing values of median \dz.
This trend breaks for the 0.1\% case, however, because this choice of target threshold results in the sampling of a very small portion of the domain.
As a result, a threshold of 0.1\% is unreliable for practical purposes and hence not considered.
Amongst the remaining choices, it is observed that selecting the top 1\% consistently results in the largest median \dz\ values for both metrics, barring regions with $\delta z<\SI{8}{\metre}$ for $-\nabla_H \cdot \mathbf{u}$ in the turbulence resolving model.
The difference in median \dz\ is as large as 100\% greater than if the threshold is decreased from 1\% to 5\% for both metrics in the operational model of the western Mediterranean.
The trend of increasing performance with increasing target threshold, given that the corresponding targets sample a sufficiently large portion of the domain, is indicative of the efficacy of the two metrics since the strongest features of any good predictor should ideally correspond to the strongest features in the quantity being predicted.

\newpage
\noindent\textbf{Text S3. Vertical displacement, surface metrics and target zones for the fully turbulent flow}

\begin{figure}[h]
    \centering
    \includegraphics[width=\linewidth]{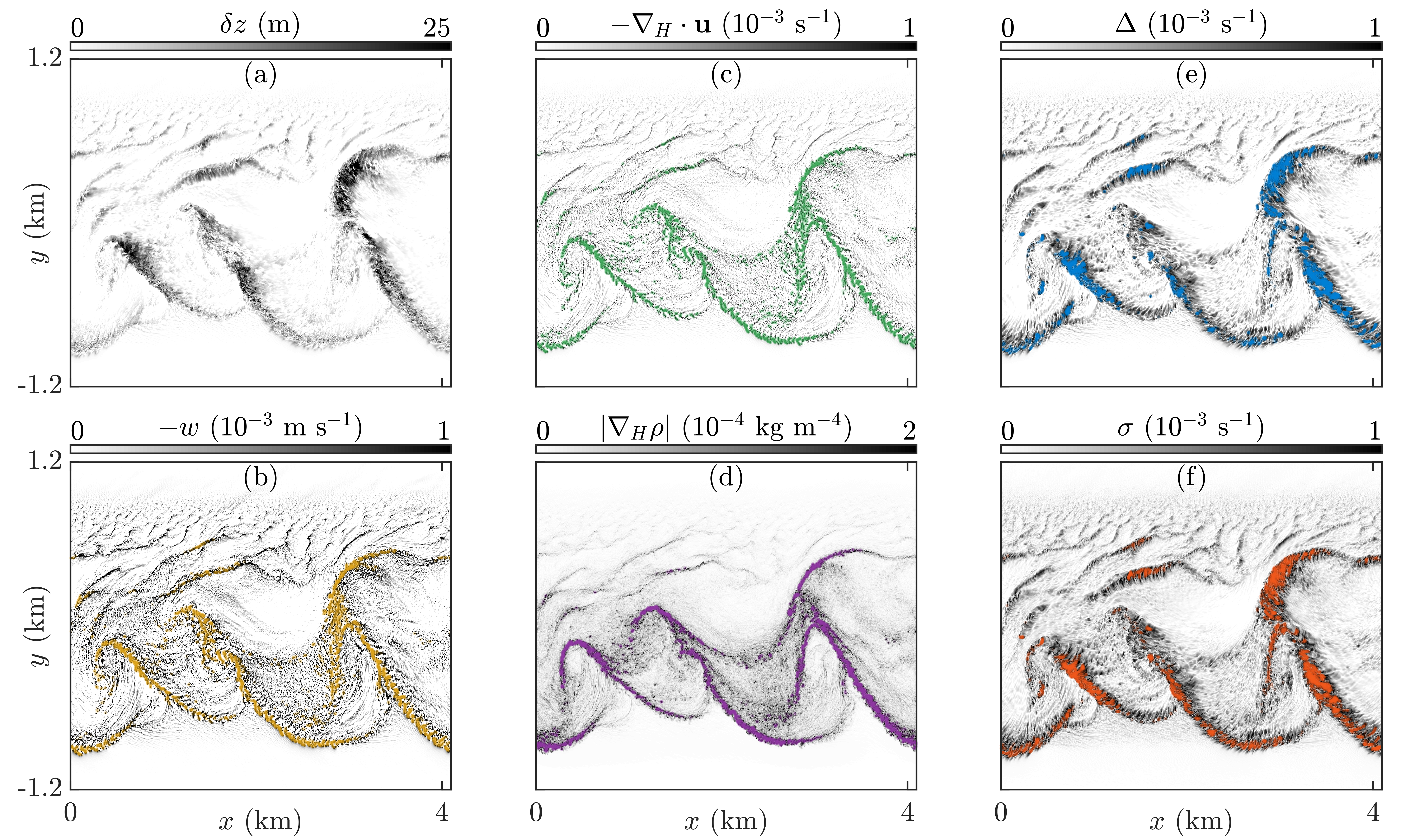}
    \caption[Sub-surface vertical displacement, surface metrics and corresponding target zones for the fully turbulent flow.]{Sub-surface vertical displacement, surface metrics and corresponding target zones for the fully turbulent flow, obtained using the turbulence resolving model. (a) Vertical displacement (\dz) field for particles released at a depth of \SI{4}{\metre}. Eulerian surface target zones: (b) negative vertical velocity ($-w$), (c) magnitude of horizontal density gradient ($|\nabla_H \rho|$), and (d) negative horizontal divergence of velocity ($-\nabla_H \cdot \mathbf{u}$). Lagrangian surface target zones: (e) dilation rate ($\Delta$) and (f) FTLE ($\sigma$). The target zones for each metric are overlaid in black in panels (b)-(f).}
    \label{fig:S1}
\end{figure}

The analysis of the turbulence resolving model presented in the main text was performed on the submesoscale fields obtained using a Lanczos filter on the fully turbulent flow fields~\cite{verma19}.
The cutoff wavenumber used for the Lanczos filter corresponds to a length scale of $O(100)$ m.
This was done since submesoscale fluid motions play the dominant role in vertical transport, as opposed to turbulent fluctuations.
Supplementary Figure~\ref{fig:S1} presents the \dz\ field, the five metrics and corresponding target zones computed using the fully turbulent fields.
The target zones are defined as the strongest regions of the metric field, with values in the 99th percentile.
The time interval of analysis was restricted to $T_d = T_L = \SI{2}{\hour}$ for this analysis due to the accumulation of numerical error during longer two-dimensional trajectory advection for the Lagrangian metric computation.

Supplementary Figure~\ref{fig:S1}a presents the \dz\ field computed over a $\SI{2}{\hour}$ interval for particles released at a depth of $\SI{4}{\metre}$.
The instantaneous Eulerian metrics calculated on the ocean surface at time $t_0$ are presented in Supplementary Figs.~\ref{fig:S1}b-d.
Regions with the largest magnitudes are near the vortex filaments, similar to \dz\ in Supplementary Fig.~\ref{fig:S1}a, but they form sharper features.
This is due to the Eulerian target zones highlighting the most dynamic regions on the surface at $t_0$.
In contrast, the \dz\ field is computed over a time interval, which allows more of the fluid to collect towards and move down the transport pathways, ultimately resulting in broader regions achieving greater \dz\.
The dilation rate and FTLE fields computed on the surface for $T_L = T_d = \SI{2}{\hour}$ are presented in Supplementary Figs.~\ref{fig:S1}e,f. 
These fields have features that are not as sharp as the Eulerian target zones and are more representative of the \dz\ field.
The target zones corresponding to each metric are overlaid in black over the respective fields in Supplementary Figs.~\ref{fig:S1}b-f. 
Even for such a short time-interval, the benefit of using Lagrangian target zones (Supplementary Figs.~\ref{fig:S1}e,f)  is apparent from their better correspondence with the high-\dz\ regions (Supplementary Fig.~\ref{fig:S1}a), compared to the Eulerian target zones (Supplementary Figs.~\ref{fig:S1}b-d).

\newpage
\noindent\textbf{Text S4. Variation of \dz\ with depth for the turbulence resolving model}

\begin{figure}[h]
    \centering
    \includegraphics[width=\linewidth]{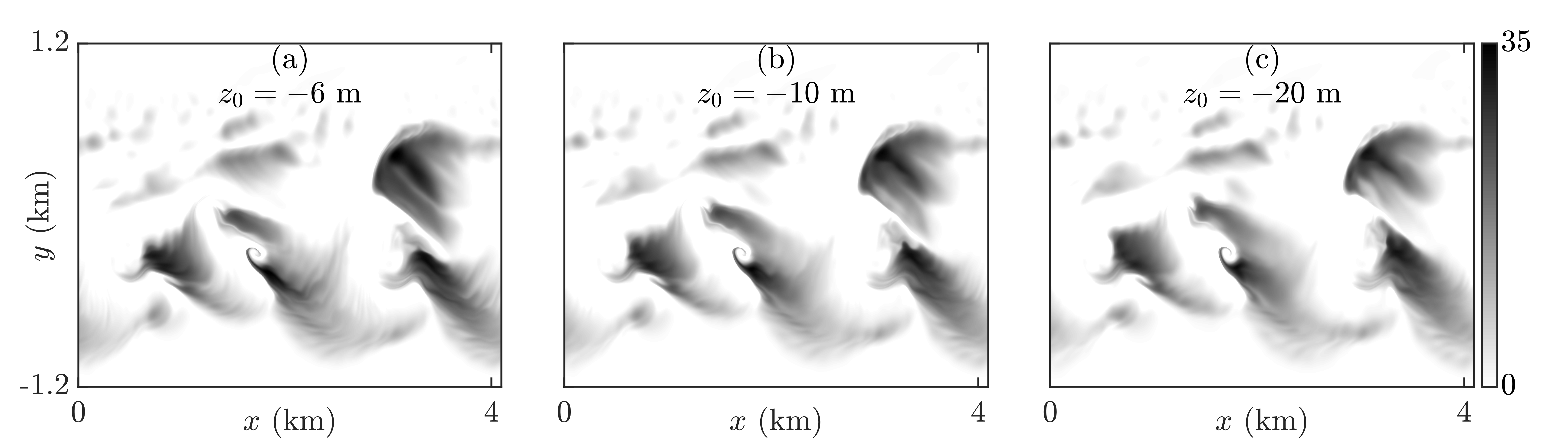}
    \caption[Variation of \dz\ field with depth for the turbulence resolving model.]{Variation of \dz field with depth for the turbulence resolving model. The \dz\ fields for particles released at depths (a) $z_0 = \SI{-6}{\metre}$, (b) $z_0 = \SI{-10}{\metre}$ and (c) $z_0 = \SI{-20}{\metre}$ are qualitatively similar to the \dz\ field presented in Fig. 1a in the main text.}
    \label{fig:S4}
\end{figure}

The analysis performed in the main text considered the \dz field at a depth of $\SI{4}{\metre}$ for the turbulence resolving model.
The choice of depth was made based on the resolution of the model, and the \dz\ field at $z_0=-\SI{4}{\metre}$ was compared to target zones derived from various Eulerian and Lagrangian metrics computed on the ocean surface.
One could argue that dynamics near the surface could be captured very well by a surface analysis, and hence, for practical applications, it is important to understand the depth to which the correspondence between \dz\ fields and surface target zones hold.

Supplementary Fig.~\ref{fig:S4}a-c present the \dz\ fields for depths of \SIlist{6;10;20}{\metre} for the turbulence resolving model. 
Strongest vertical displacement occurs at the same locations for all three depths, and these regions are reminiscent of the high-\dz\ regions for the \SI{4}{\metre} case presented in Fig.~2a.
This indicates that surface target zones that identify the strongest \dz\ regions for a depth of \SI{4}{\metre} can successfully identify such regions even for a depth of \SI{20}{\metre} for the turbulence resolving model.
The magnitudes of \dz\ also remain similar with depth, indicating that median \dz\ experienced by particles released within target zones at various depths will be similar.

\begin{table}[ht]
\centering
\begin{tabular}{ |c|c|c|c|c|c| } 
    \hline
    $z_0$ (m) & -6 & -8 & -10 & -20 & -30 \\ 
    \hline
    Turbulence resolving model & 0.97 & 0.90 & 0.80 & 0.55 & 0.41 \\ 
    \hline
    Process study model & 0.99 & 0.97 & 0.94 & 0.80 & 0.67 \\ 
    \hline
    Operational model & 0.97 & 0.91 & 0.87 & 0.69 & 0.54 \\ 
    \hline
\end{tabular}
\caption[Variation of \dz\ fields with depth for the three models.]{Variation of \dz\ fields with depth for the three models. Correlation coefficients ($c_{\delta z}$) between \dz\ fields at various depths and the \dz\ field at $z_0 = \SI{-4}{\metre}$ are tabulated for all three models.}
\label{tab:S1}
\end{table}

To investigate the degree of similarity of \dz\ fields with depth for the three models, we present the correlation coefficient ($c_{\delta z}$) between the \dz\ field at \SI{4}{\metre} depth and the fields at \SIlist{6;8;10;20;30}{\metre} in Supplementary Table~\ref{tab:S1}.
For the turbulence resolving model, the value of the correlation coefficient decays quickly with depth because of the presence of strong three-dimensional motions.
However, even for a depth of \SI{20}{\metre} where $c_{\delta z} = 0.55$, the strongest vertical displacement zones are comparable to the high-\dz\ regions at $\SI{4}{\metre}$, as indicated by Supplementary Fig.~\ref{fig:S4}c.
Similar to the turbulence resolving model, the \dz\ fields at \SI{10}{\metre} for the process study and operational models are highly correlated ($c_{\delta z} \approx 0.9$) to the respective fields at \SI{4}{\metre}.
The correlation coefficients for these models do not drop below $0.5$ even for a depth \SI{30}{\metre}, indicating that our surface target zones are likely to capture strong vertical displacement zones at deeper layers.

\newpage
\noindent\textbf{Text S5. Target zone overlays on \dz\ fields}

\begin{figure}[h]
    \centering
    \includegraphics[width=\linewidth]{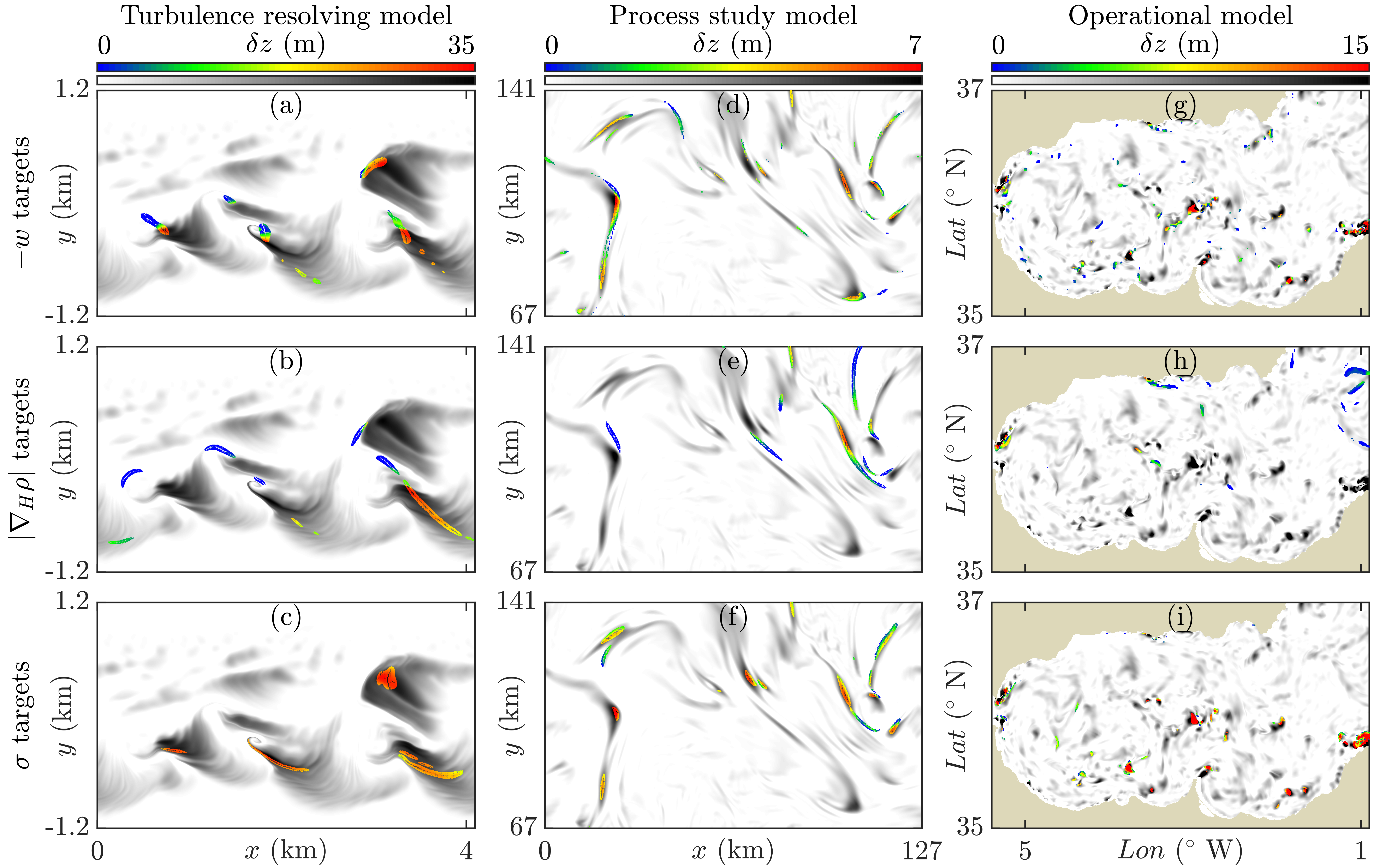}
    \caption[Target zone overlays on \dz\ fields.]{Target zone overlays on \dz\ fields. Target zones for the metrics $-w$ (top), $|\nabla_H \rho|$ (middle) and $\sigma$ (bottom), overlaid using a colormap that is indicative of the \dz\ values plotted underneath in grayscale.
    Shades of blue correspond to regions where the target has missed a vertical displacement conduit and red regions represent target zones that have identified the strongest vertical displacement conduits.
    Results are presented for the turbulence resolving model (left), process study model (centre) and operational model of the western Mediterranean (right).}
    \label{fig:S5}
\end{figure}

A qualitative understanding of the ability of each surface target zone to identify strong vertical displacement zones below the surface is obtained from overlays of target zones on the \dz\ field.
Figure~3 in the main text presents such overlays for an Eulerian metric ($-\nabla_H \cdot \mathbf{u}$) and a Lagrangian metric ($\Delta$), and it was observed that the $\Delta$ targets were better at identifying strong vertical displacement zones.
Supplementary Figure~\ref{fig:S5} presents similar overlays for the three other metrics considered: $-w$, $|\nabla_H \rho|$ and $\sigma$.

For the turbulence resolving and operational models, the $-w$ targets are similar to the $-\nabla_H \cdot \mathbf{u}$ targets, but it is not the case for the process study model.
The $|\nabla_H \rho|$ targets are qualitatively different from the other Eulerian target zones for all three models, while also failing to capture strong vertical displacement zones.
This indicates that while $|\nabla_H \rho|$ reveals the surface signatures of density fronts that are sites of intense vertical motions, regions where the metric attains its largest values are not necessarily linked to strongest vertical displacement.
The Lagrangian target zone, on the other hand, successfully identifies significant portions of high-\dz\, as evidenced by the near-absence of blue regions in Supplementary Figs.~\ref{fig:S5}c,f,i.
It is noteworthy that the $\sigma$ targets for the turbulence resolving model (Supplementary Fig.~\ref{fig:S5}c) captures a fourth high-\dz\ region compared to the $\Delta$ targets in Fig.~3b.

\newpage
\noindent\textbf{Text S6. Ensemble analysis of the operational model}

\begin{figure}[h]
    \centering
    \includegraphics[width=\linewidth]{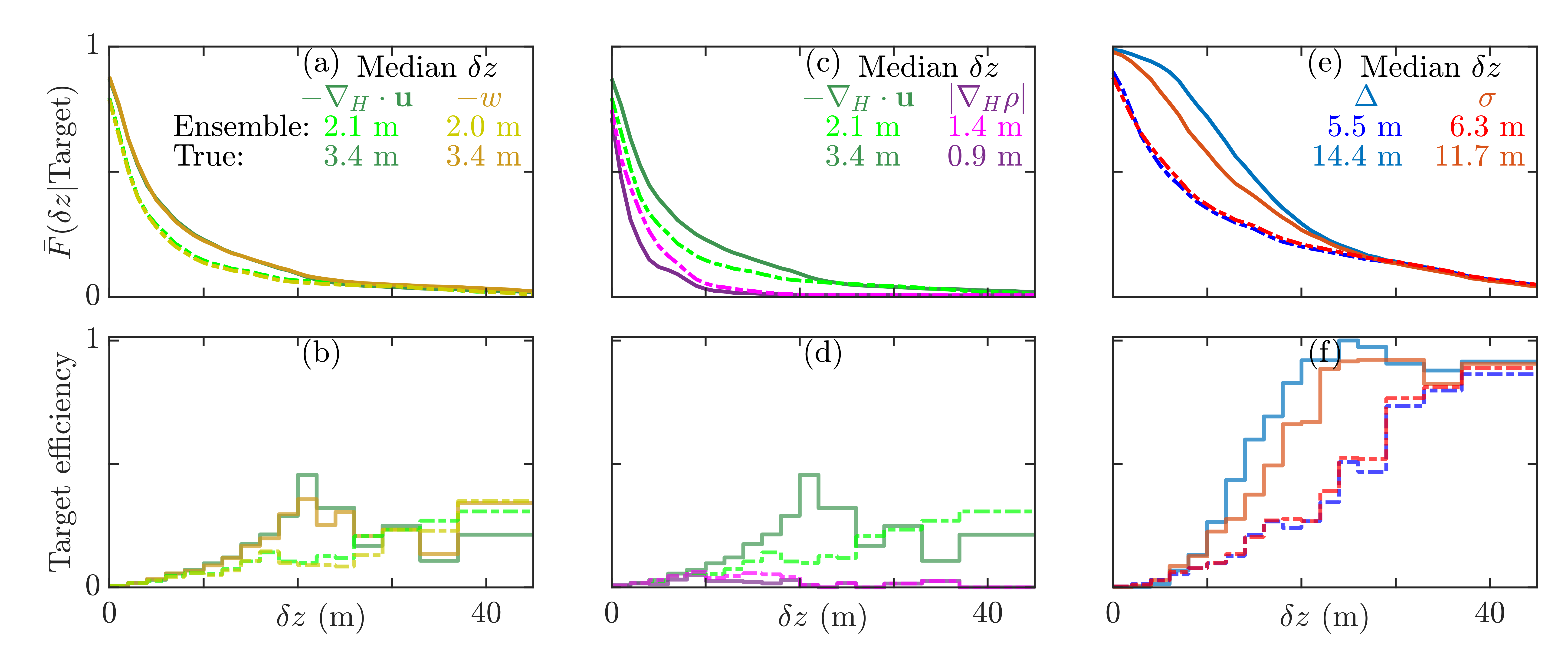}
    \caption[Ensemble analysis of the operational model.]{Ensemble analysis of the operational model. Complementary CDFs (top) and target efficiency (bottom) for the ensemble target zones (dashed lines) and true target zones (solid lines) that are not presented in Fig.~6e,f. Skill curves are plotted as a function of \dz\ for the target zones of Eulerian metrics $-w$ (left) and $|\nabla_H \rho|$ (centre), and the Lagrangian metric $\sigma$ (right). Corresponding curves for the $-\nabla_H \cdot \mathbf{u}$ target zones (Eulerian) and the $\Delta$ target zones (Lagrangian) are plotted for comparison wherever required. Median \dz\ for each metric is also indicated.}
    \label{fig:S6}
\end{figure}

For the ensemble analysis in the main text, Fig.~6 presents results only for one Eulerian ($-\nabla_H \cdot \mathbf{u}$) and Lagrangian ($\Delta$) target zones.
Supplementary Figure~\ref{fig:S6} presents the complementary CDFs and the target efficiency for the true and ensemble target zones obtained using the remaining metrics.

The skill curves from Fig.~6e,f are also presented for comparison, and it is observed that both the true and ensemble $-w$ target zones are comparable to $-\nabla_H \cdot \mathbf{u}$.
The $|\nabla_H \rho|$ target zones perform worse than the other Eulerian target zones, and interestingly, the ensemble target zones performs better than the true target zones.
This highlights the sensitivity of the $|\nabla_H \rho|$ target zone to model uncertainty, and indicates that density gradient is not a good choice.
The Lagrangian $\sigma$ target zone, on the other hand, performs better than all three Eulerian target zones.
The true and ensemble $\sigma$ target zones, however, perform slightly worse than $\Delta$, especially for \dz$<\SI{30}{\metre}$.

\newpage
\bibliographystyle{elsarticle-harv} 
\bibliography{references}